\documentclass[aps,showpacs,nofootinbib,superscriptaddress]{revtex4}
\usepackage{graphicx}
\usepackage{dcolumn}

\def\slashchar#1{\setbox0=\hbox{$#1$}
   \dimen0=\wd0 \setbox1=\hbox{/} \dimen1=\wd1
   \ifdim\dimen0>\dimen1 \rlap{\hbox to \dimen0{\hfil/\hfil}} #1
   \else  \rlap{\hbox to \dimen1{\hfil$#1$\hfil}} / \fi}
\begin{document}
\title{Inclusive Charged--Current Neutrino--Nucleus Reactions}

\author{J. Nieves}
\affiliation{Instituto de F\'\i sica Corpuscular (IFIC), Centro Mixto
Universidad de Valencia-CSIC, Institutos de Investigaci\'on de
Paterna, E-46071 Valencia, Spain}
\author{I. \surname{Ruiz Simo}}
\affiliation{Departamento de F\'\i sica Te\'orica and IFIC, Centro Mixto
Universidad de Valencia-CSIC, Institutos de Investigaci\'on de
Paterna, E-46071 Valencia, Spain}
\author{M. J.  \surname{Vicente Vacas}}
\affiliation{Departamento de F\'\i sica Te\'orica and IFIC, Centro Mixto
Universidad de Valencia-CSIC, Institutos de Investigaci\'on de
Paterna, E-46071 Valencia, Spain}

\today

\begin{abstract}

We present  a model for weak CC induced
nuclear reactions at energies of interest for current
and future neutrino oscillation experiments. This model is a natural
extension of the work of Refs.~\cite{Gil:1997bm, Nieves:2004wx}, where
the QE contribution to the inclusive electron and neutrino scattering on
nuclei was analyzed.  The model is
based on a systematic many body expansion of the gauge boson
absorption modes that includes one, two and even three body
mechanisms, as well as the excitation of $\Delta$ isobars. The whole
scheme has no free parameters, besides those previously adjusted to
the weak pion production off the nucleon cross sections in the
deuteron, since all nuclear effects were set up in previous studies of
photon, electron and pion interactions with nuclei. We have discussed
at length the recent charged current quasi-elastic MiniBooNE cross
section data, and showed that two nucleon knockout mechanisms are
essential to describe these measurements.

\end{abstract}


\maketitle

\section{Introduction}

The interaction of neutrinos with nuclei at intermediate energies
plays an important role in the precise determination of neutrino
properties such as their masses and mixing parameters. It can also
provide relevant information on the axial hadronic currents. The
statistical significance of the experiments is rapidly improving.
However, the data analysis needs to consider a large number of nuclear
effects that distort the signals and produce new sources of background
that are absent in the elementary neutrino nucleon processes.

In this context, it is clearly of interest the elaboration of a
theoretically well founded and unified framework in which the
electroweak interactions with nuclei could be systematically
studied. Furthermore, the recent measurements of the cross sections
for several
channels~\cite{AguilarArevalo:2010zc,AguilarArevalo:2010cx,AguilarArevalo:2010bm,Nakajima:2010fp}
provide a serious benchmark to the theoretical models. An excellent
review of the current situation can be found in
Ref.~\cite{AlvarezRuso:2010ia}.

A suitable theoretical model should include, at least, three kinds of
contributions: (i) quasielastic (QE) for low energy transfers, (ii)
pion production and two-body processes from the QE region to that
around the $\Delta(1232)$ resonance peak, and (iii) double pion
production and higher nucleon resonance degrees of freedom induced
processes at even higher energies. A word of caution is needed here,
because the same words could refer to somehow different magnitudes in
the literature. For instance, whereas in most theoretical works QE is
used for processes where the gauge boson $W^{\pm}$ or $Z^0$ is
absorbed by just one nucleon, which together with a lepton is
emitted\footnote{This follows the traditional nomenclature of
electronuclear scattering. Note that in same cases the resulting
nucleon after the absorption of the gauge boson is not emitted, but
rather it could be trapped and form part of a bound state of the
daughter nucleus (discrete transition).}, in the recent MiniBooNE
papers, QE is related to processes in which only a muon is
detected. This latter definition could make sense because ejected
nucleons are not detected in that experiment, but includes
multinucleon processes and others like pion production followed by
absorption. However, it discards pions coming off the nucleus, since
they will give rise to additional leptons after their decay. In any
case, their experimental results cannot be directly compared to most
previous calculations.

The QE processes have been abundantly studied. Simple approaches using
a global Fermi gas for the nucleons and the impulse approximation are
good enough to describe qualitatively electron scattering but more
sophisticated treatments of the nuclear effects are necessary to get a
detailed agreement with data. There are different kinds of models like
those based on the use of proper nucleon spectral
functions~\cite{Benhar:2005dj,Ankowski:2007uy,Ankowski:2010yh}, others in which
nucleons are treated in a relativistic mean
field~\cite{Maieron:2003df, Martinez:2005xe} and models based on a
local Fermi gas including many body effects such as spectral
functions~\cite{Leitner:2008ue} and
RPA~\cite{Nieves:2004wx,Nieves:2005rq,
Singh:1992dc,SajjadAthar:2009rd}. Concerning the elementary process,
$\nu+ N\rightarrow l+N'$, the hadronic vector current is well known
from electron scattering. The axial current, after the use of the
partial conservation of the axial current to relate the two form
factors and assuming a dipole form, depends on two parameters: $g_A$,
that can be fixed from the neutron $\beta$ decay and the axial mass $M_A$. The
value of $M_A$ established from QE data on deuterium targets is
$M_A=1.016 \pm 0.026$~\cite{Bodek:2007ym} GeV. A consistent result is
obtained from $\pi$ electro-production after chiral corrections are
incorporated~\cite{Liesenfeld:1999mv,Bernard:1992ys}.

The predicted cross sections for QE scattering are very similar for
most models. See, e.g., the compilation shown in Fig. 2 of
Ref.~\cite{Boyd:2009zz}. On the other hand, the theoretical results
are clearly below the recently published MiniBooNE
data~\cite{AguilarArevalo:2010zc}. The discrepancy is large enough to
provoke much debate and theoretical attention. Some works try to
understand these new data in terms of a larger value of $M_A$. For
instance, in Ref.~\cite{AguilarArevalo:2010zc} a value of $M_A=1.35\pm
0.17$, that also fits the $Q^2$ shape, is suggested. Consistent values
are obtained in
Refs.~\cite{Butkevich:2010cr,Benhar:2010nx,Juszczak:2010ve}. This idea
is not only difficult to understand theoretically, but is also in
conflict with higher energy NOMAD data~\cite{Lyubushkin:2008pe}
($M_A=1.06 \pm 0.02(stat) \pm 0.06(syst)$ GeV). In another line of
research, the role of meson exchange currents~\cite{Amaro:2010sd} and
superscaling~\cite{Amaro:2010qn} have  been also estimated
recently. Finally, another idea has been explored in
Refs.~\cite{Martini:2009uj,Martini:2010ex}, which include two nucleon
mechanisms (and others related to $\Delta$ excitation) and reproduce
MiniBooNE QE data without the need of a large value of $M_A$.  These
latter results suggest that much of the experimental cross section can
be attributed to processes that are not properly QE, stressing again
the need of a unified framework dealing with all relevant mechanisms,
namely $\pi$ production and multinucleon excitation.

The matter of $\pi$ production induced by neutrinos is also of much
interest~\cite{AlvarezRuso:1998hi,Sato:2003rq,Hernandez:2007qq,Leitner:2008wx,Leitner:2010jv,Graczyk:2009qm,Hernandez:2010bx,Lalakulich:2010ss}. The
elementary reaction on the nucleon, at low and intermediate energies,
includes both background and resonant mechanisms. The background terms
can be obtained from the chiral lagrangians. The resonant terms
contain some free parameters that have been adjusted to ANL and/or BNL
old bubble chamber data. Still, the experimental data have large
normalization uncertainties which are certainly reflected in the
theoretical models. At low energies, the $\Delta(1232)$ resonance
plays a very important role in this process, and for small $Q^2$
values only one form factor ($C_5^A$) is relevant. Thus, special
attention has been addressed to its study with recent results ranging
from $C_5^A(0)=1.19\pm 0.08$~\cite{Graczyk:2009qm}, obtained
neglecting the non resonant background, to $C_5^A(0)=1.00\pm
0.11$~\cite{Hernandez:2010bx} in a more complete model. This latter
value implies a 20\% reduction with respect to the off--diagonal
Goldberger-Treiman relation (GTR).  In nuclei, several effects are expected
to be important for the $\pi$ production reaction. First, the
elementary process is modified by Fermi motion, by Pauli blocking and
more importantly by the changes of the spectral function of the
$\Delta$ resonance in the medium. In addition, the final pion can be
absorbed or scattered by one or more nucleons.  This latter kind of
effects do not modify the inclusive neutrino nucleus cross section and
thus are out of the scope of this paper.

Our aim in this work is to extend the model of
Ref.~\cite{Nieves:2004wx}, which studied QE scattering. We will
include two nucleon processes and $\pi$ production in a well
established framework that has been tested, for instance, in electron
and photon scattering~\cite{Carrasco:1989vq, Gil:1997bm}. This will
extend the range of applicability of the model to higher transferred
energies (and thus higher neutrino energies) and allow for the
comparison with inclusive data which include the QE peak, the $\Delta$
resonance peak and also the dip region between them.  The structure of
the paper is as follows: In Sect.~\ref{sec:neu}, we start establishing
the formalism and reviewing briefly the approach for QE scattering of
Ref.~\cite{Nieves:2004wx}. Then, we consider pion production
mechanisms and two nucleon processes. Next, we discuss with special
care the role of the $\Delta$ resonance and how it is affected by the
nuclear medium. In Sect.~\ref{sec:res} we present and discuss some of
the results derived from the model, and compare these to the recent
MiniBooNE charged current (CC) QE and SciBooNE total cross section
data. Finally in Sect.~\ref{sec:conc}, we draw the main conclusions of
this work.

\section{CC  neutrino/antineutrino inclusive nuclear reactions} 
\label{sec:neu}
\subsection{General considerations}

We will focus on the inclusive nuclear reaction 
\begin{equation}
\nu_l (k) +\, A_Z \to l^- (k^\prime) + X 
\label{eq:reac}
\end{equation}
driven by the electroweak CC. The generalization of the obtained
expressions to antineutrino induced reactions is straightforward (see
Subsect.~\ref{sec:anti} ). The double differential cross section, with
respect to the outgoing lepton kinematical variables, for the process
of Eq.~(\ref{eq:reac}) is given in the Laboratory (LAB) frame by
\begin{equation}
\frac{d^2\sigma_{\nu l}}{d\Omega(\hat{k^\prime})dE^\prime_l} =
\frac{|\vec{k}^\prime|}{|\vec{k}~|}\frac{G^2}{4\pi^2} 
L_{\mu\sigma}W^{\mu\sigma} \label{eq:sec}
\end{equation}
with $\vec{k}$ and $\vec{k}^\prime~$ the LAB lepton momenta, $E^{\prime}_l =
(\vec{k}^{\prime\, 2} + m_l^2 )^{1/2}$ and $m_l$ the energy and the
mass of the outgoing lepton, $G=1.1664\times 10^{-11}$ MeV$^{-2}$, the
Fermi constant and $L$ and $W$ the leptonic and hadronic tensors,
respectively. To obtain Eq.~(\ref{eq:sec}) we have
neglected the four-momentum carried out by the intermediate $W-$boson
with respect to its mass, and have used  the  relation between
the gauge weak coupling
constant, $g = e/\sin \theta_W$, and  the Fermi constant:  
$G/\sqrt 2 = g^2/8M^2_W$, with $e$ the electron charge, $\theta_W$ the
Weinberg angle and $M_W$ the $W-$boson mass.
 The leptonic tensor is given by\footnote{We
take $\epsilon_{0123}= +1$ and the metric $g^{\mu\nu}=(+,-,-,-)$.}:
\begin{eqnarray}
L_{\mu\sigma}&=& L^s_{\mu\sigma}+ {\rm i} L^a_{\mu\sigma} =
 k^\prime_\mu k_\sigma +k^\prime_\sigma k_\mu
- g_{\mu\sigma} k\cdot k^\prime + {\rm i}
\epsilon_{\mu\sigma\alpha\beta}k^{\prime\alpha}k^\beta \label{eq:lep}.
\end{eqnarray}
The hadronic tensor corresponds to the charged electroweak transitions of the
target nucleus, $i$, to all possible final states. 
It is  given by\footnote{In 
Eq.~(\ref{eq:wmunu}) the states are normalized such that $\langle
\vec{p} | \vec{p}^{\,\prime} \rangle = (2\pi)^3 2p_0
\delta^3(\vec{p}-\vec{p}^{\,\prime})$ and the sum over final states
$f$ includes an integration $ \int \frac{d^3p_j}{(2\pi)^3 2E_j}$, for
each particle $j$ making up the system $f$, as well as a sum over all
spins involved.}
\begin{eqnarray}
W^{\mu\sigma} &=& \frac{1}{2M_i}\overline{\sum_f } (2\pi)^3
\delta^4(P^\prime_f-P-q) \langle f | j^\mu_{\rm cc}(0) | i \rangle
 \langle f | j^\sigma_{\rm cc}(0) | i \rangle^*
\label{eq:wmunu}
\end{eqnarray}
with $P$ the four-momentum of the initial nucleus, $M_i=\sqrt{P^2}$ 
the target nucleus mass, $P_f^\prime$  the total four momentum of
the hadronic state $f$ and $q=k-k^\prime$ the four momentum
transferred to the nucleus.  The bar over the sum denotes the
average over initial spins. The hadronic CC  is given by
\begin{equation}
j^\mu_{\rm cc} = \overline{\Psi}_u\gamma^\mu(1-\gamma_5)(\cos\theta_C \Psi_d +
\sin\theta_C \Psi_s) 
\end{equation}
with $\Psi_u$, $\Psi_d$ and $\Psi_s$ quark fields, and $\theta_C$ the
Cabibbo angle. By construction, the hadronic
tensor accomplishes
\begin{eqnarray}
W^{\mu\sigma}= W^{\mu\sigma}_s + {\rm i} W^{\mu\sigma}_a 
\end{eqnarray}
with $W^{\mu\sigma}_s$ ($W^{\mu\sigma}_a$) real symmetric
(antisymmetric) tensors. The hadronic tensor is determined by the
$W^+-$boson selfenergy, $\Pi^{\mu\sigma}_W(q)$, in the nuclear medium.
We follow here the formalism of Ref.~\cite{Nieves:2004wx}, where it is
shown that within the local density approximation, the hadronic tensor
can be written as
\begin{eqnarray}
W^{\mu\sigma}_s &=& - \Theta(q^0) \left (\frac{2\sqrt 2}{g} \right )^2 
\int \frac{d^3 r}{2\pi}~ {\rm Im}\left [ \Pi_W^{\mu\sigma} 
+ \Pi_W^{\sigma\mu} \right ] (q;\rho)\label{eq:wmunus}\\
W^{\mu\sigma}_a &=& - \Theta(q^0) \left (\frac{2\sqrt 2}{g} \right )^2
 \int \frac{d^3 r}{2\pi}~{\rm Re}\left [ \Pi_W^{\mu\sigma} 
- \Pi_W^{\sigma\mu}\right] (q;\rho) \label{eq:wmunua}.
\end{eqnarray}
Then, the
differential cross section for the reaction in Eq.~(\ref{eq:reac})
is given by 
\begin{eqnarray}
\frac{d^2\sigma_{\nu l}}{d\Omega(\hat{k^\prime})dk^{\prime 0}} &=&-
\frac{|\vec{k}^\prime|}{|\vec{k}~|}\frac{G^2}{2\pi^2}
\left(\frac{2\sqrt 2}{g}\right)^2
 \int\frac{d^{\,3}r}{2\pi} ~{\rm
  Im} \Big [  L_{\mu\eta} \Pi_W^{\eta\mu}(q;\rho) \Big ] \Theta(q^0) \nonumber
 \\
&=& -
\frac{|\vec{k}^\prime|}{|\vec{k}~|}\frac{G^2}{4\pi^2}
\left(\frac{2\sqrt 2}{g}\right)^2
 \int\frac{d^{\,3}r}{2\pi} \Big \{   L_{\mu\eta}^s ~{\rm
  Im}\left[\Pi_W^{\mu\eta}+\Pi_W^{\eta\mu}\right] 
 -     L_{\mu\eta}^a ~{\rm
  Re}\left[\Pi_W^{\mu\eta}-\Pi_W^{\eta\mu}\right] \Big \} \Theta(q^0)
\end{eqnarray}
with $\Theta(...)$ the Heaviside step function.

\begin{figure}[tbh]
\centerline{\includegraphics[height=9.0cm]{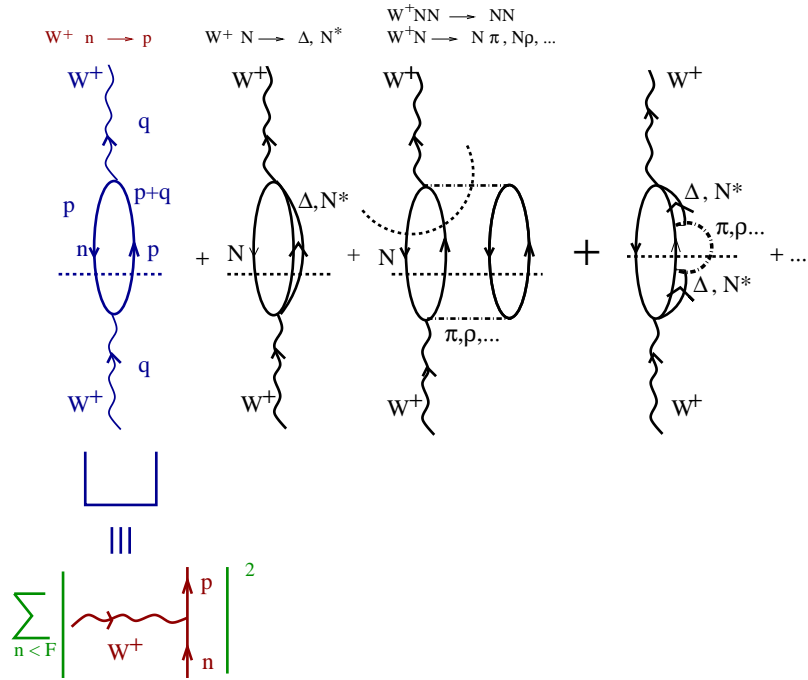}}
\caption{\footnotesize Diagrammatic representation of some mechanisms
  contributing to the $W^+-$selfenergy. }\label{fig:fig1}
\end{figure}

The in medium gauge boson ($W^{\pm}$) selfenergy depends on the
nuclear density $\rho(r)$. We propose  a many body expansion for
$\Pi_W^{\mu\sigma}$, where the relevant gauge boson absorption modes
would be  systematically incorporated: absorption by one nucleon, or a
pair of nucleons or even three nucleon mechanisms, real and virtual
meson ($\pi$, $\rho$, $\cdots$) production, excitation of $\Delta$ of
higher resonance degrees of freedom, etc. In addition, nuclear effects
such as RPA or Short Range Correlations (SRC) will be
also taken into account.  Some of the basic $W-$absorption modes are
depicted in Fig.~\ref{fig:fig1}.

\subsection{Quasielastic scattering}
The virtual $W^+$ can be absorbed by one nucleon leading to the QE
contribution of the nuclear response function. Such a process
corresponds to a one particle-one hole (1p1h) nuclear excitation
(first of the diagrams depicted in Fig.~\ref{fig:fig1}).  This
contribution was studied in detail in
Ref.~\cite{Nieves:2004wx}\footnote{The extension of this scheme for
Neutral Currents (NC) was discussed in
Ref.~\cite{Nieves:2005rq}.}. Here, we will just briefly discuss the
main features of the model.  In Ref.~\cite{Nieves:2004wx}, starting
from a Local Fermi Gas (LFG) picture of the nucleus, which
automatically accounts for Pauli blocking and Fermi motion, several
nuclear corrections were incorporated, among others:
\begin{itemize}
\item A correct
energy balance, using the experimental $Q-$values, was enforced.
\item  Coulomb distortion of the charged leptons, important at low
  energies,  was implemented by using
the so called ``modified effective momentum approximation''.  
\item Medium polarization (RPA), including $\Delta-$hole degrees of
freedom and explicit pion and rho exchanges in the vector--isovector
channel of the effective nucleon--nucleon force, and SRC effects were
 computed.
\item The nucleon propagators were dressed in the nuclear medium,
which amounts to work with nucleon spectral functions (a LFG of
interacting nucleons) and it also accounts for some reaction
mechanisms where the gauge boson is absorbed by two nucleons.
\end{itemize}
This model is a natural extension of previous studies on
electron~\cite{Gil:1997bm},
photon~\cite{Carrasco:1989vq} and
pion~\cite{Oset:1981ih,Salcedo:1987md, Nieves:1993ev, Nieves:1991ye,
  Albertus:2001pb}
dynamics in nuclei. Even though the scarce existing CC data involve
very low nuclear excitation energies, for which specific details of
the nuclear structure might play an important role, the model of
Ref.~\cite{Nieves:2004wx} provides one of the best existing combined
descriptions of the inclusive muon capture in $^{12}$C and of the
$^{12}$C $(\nu_\mu,\mu^-)X$ and $^{12}$C $(\nu_e,e^-)X$ reactions near
threshold.  Inclusive muon capture from other nuclei is also
successfully described. 

The theoretical errors affecting the predictions of Ref.~\cite{Nieves:2004wx}
were  discussed in Ref.~\cite{Valverde:2006zn}. There, it is concluded
that is sound to assume errors of about 10-15\% on the QE
neutrino--nucleus (differential and integrated) cross section results
of Ref.~\cite{Nieves:2004wx}.

The LFG description of the nucleus has been shown to be well suited
for inclusive processes and nuclear excitation energies of around 100
MeV or higher~\cite{Gil:1997bm, Carrasco:1989vq,
Oset:1981ih,Salcedo:1987md, Nieves:1993ev, Nieves:1991ye,
Albertus:2001pb}. The reason is that in these circumstances one should
sum up over several nuclear configurations, both in the discrete and
in the continuum. This inclusive sum is almost insensitive to the
details of the nuclear wave function\footnote{The results of
Ref.~\cite{Nieves:2004wx} for the inclusive muon capture in nuclei
through the whole periodic table, where the capture widths vary from
about 4$\times 10^4$ s$^{-1}$ in $^{12}$C to 1300 $\times 10^4$
s$^{-1}$ in $^{208}$Pb, and of the LSND measurements of the $^{12}$C
$(\nu_\mu,\mu^-)X$ and $^{12}$C $(\nu_e,e^-)X$ reactions near
threshold indicate that the predictions of our scheme, for totally
integrated inclusive observables, could be extended to much
smaller, ($\approx$ 10 or 20 MeV), nuclear excitation energies.  In
this respect, Refs.~\cite{Amaro:2004cm} and ~\cite{Amaro:1997ed} for
inclusive muon capture and radiative pion capture in nuclei,
respectively, are enlightening. In these works, continuum shell model
and LFG model results are compared for several nuclei from $^{12}$C to
$^{208}$Pb. The differential decay width shapes predicted for the two
models are substantially different. Shell model distributions present
discrete contributions and in the continuum appear sharp scattering
resonances. Despite the fact that those distinctive features do not
appear in the LFG differential decay widths, the totally integrated
widths (inclusive observable) obtained from both descriptions of the
process do not differ in more than 5 or 10\%. The typical nuclear
excitation energies in muon and radiative pion capture in nuclei are
small, of the order of 20 MeV, and thus one expects that at higher
excitation energies, where one should sum up over a larger number of
nuclear final states, the LFG predictions for inclusive observables
would become even more reliable.}, in sharp contrast to what happens
in the case of exclusive processes where the final nucleus is left in
a determined nuclear level. On the other hand, the LFG description of
the nucleus allows for an accurate treatment of the dynamics of the
elementary processes (interaction of gauge bosons with nucleons,
nucleon resonances, and mesons, interaction between nucleons or
between mesons and nucleons, etc.)  which occur inside the nuclear
medium. Within a finite nuclei scenario, such a treatment becomes hard
to implement, and often the dynamics is simplified in order to deal
with more elaborated nuclear wave functions.

\subsection{The virtual $W-$self-energy in pion production: 1p1h$1\pi$
  contribution}

\begin{figure}
\includegraphics[height=.28\textheight]{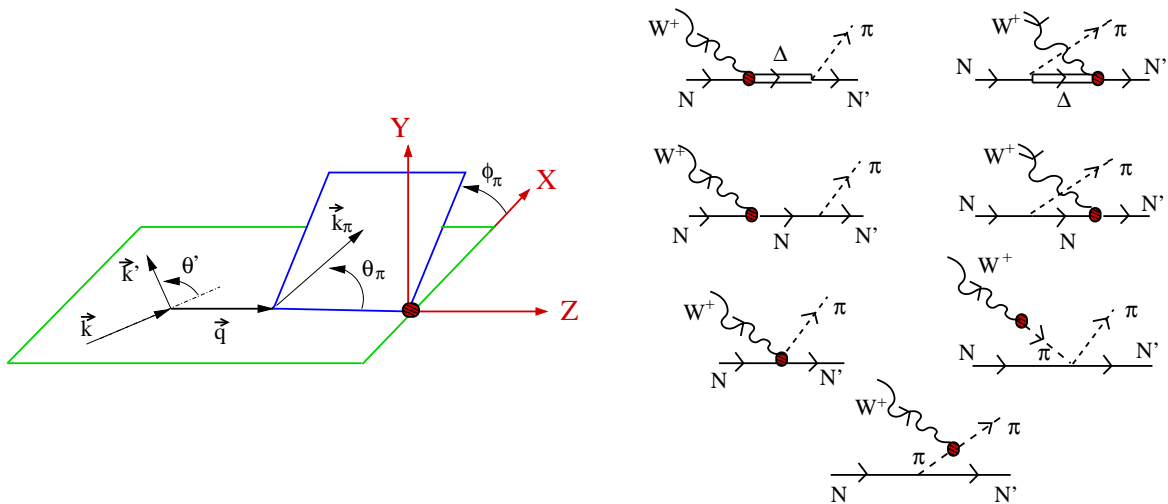}
  \caption{Left: Definition of the  kinematical variables
used in this work. Right: Model for the $W^+ N\to N^\prime\pi$
  reaction. It consists of seven diagrams: Direct and crossed
  $\Delta(1232)-$ (first row) and nucleon (second row) pole terms,
  contact and pion pole contribution (third row) and finally the
  pion-in-flight term.  Throughout this work, we will label these
  contributions by: $\Delta P$, $C\Delta P$, $NP$, $CNP$, $CT$, $PP$ and
  $PF$, respectively. The circle in the diagrams stands for the weak
  transition vertex.}
\label{fig:diagramas}
\end{figure}

In this subsection, we  calculate the contribution to the cross section
from $W^+$ gauge boson self-energy diagrams  which
contains pion production in the intermediate states. We will use 
the model for the CC neutrino--pion production reaction off the
nucleon, 
\begin{equation}
  \nu_l (k) +\, N(p)  \to l^- (k^\prime) + N(p^\prime) +\, \pi(k_\pi) 
\end{equation}
derived in Refs.~\cite{Hernandez:2007qq,Hernandez:2010bx}. This
process, at intermediate energies, is traditionally described in the
literature by means of the weak excitation of the $\Delta(1232)$
resonance and its subsequent decay into $N\pi$. In
Ref.~\cite{Hernandez:2007qq}, some background terms required by the
pattern of spontaneous chiral symmetry breaking of QCD are also
included. Their contributions are sizable and lead to significant
effects in total and partially integrated pion production cross
sections even at the $\Delta(1232)-$resonance peak, and they are
dominant near pion threshold.  The model consists of seven diagrams
(right panel of Fig.~\ref{fig:diagramas}). The  contributions of the
different diagrams are
calculated by using the effective Lagrangian of the SU(2) nonlinear
$\sigma-$model, supplemented with some form--factors (see
Ref.~\cite{Hernandez:2007qq} for details). In this work, we will use
the set IV of form factors compiled in Table I of
Ref.~\cite{Hernandez:2010bx}.  The available data set on neutrino and
antineutrino pion production on nucleons is described reasonably
well. Nonetheless, we must mention, that the experimental data still
have large uncertainties and there exist conflicting data for some
channels.

The discussed model can be considered an extension of that developed
in Ref.~\cite{Gil:1997bm} for the $ e N \to e' N \pi$ reaction.  For
the latter case, the model, that contains a theoretically well founded
description of the background amplitudes, provides the same level of
accuracy~\cite{Lalakulich:2010ss} as the MAID
model~\cite{Drechsel:2007if}, which ensures its applicability to the
leptoproduction processes at least up to $W < 1.4$ GeV, being $W$ the
outgoing $\pi N$ invariant mass.

\begin{figure}
\makebox[0pt]{\includegraphics[height=.24\textheight]{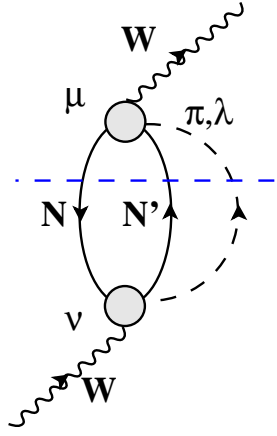}}
 \caption  {$W-$self-energy obtained by folding the 
 $W N \rightarrow \pi N'$ amplitude ($\lambda$ is the charge of the
 pion).}
\label{fig:1ph-pi}
\end{figure}

We move now to the computation of the  $W^+$ gauge boson self-energy diagrams which contain
pion production in the intermediate states. This is readily
accomplished by taking  the $W^+ N \rightarrow
\pi N'$ amplitude of Fig.~\ref{fig:diagramas} and folding it with
itself. One gets then the diagram of Fig.~\ref{fig:1ph-pi} where the
circle stands for any of the 7 terms of the elementary model for
$W N \rightarrow \pi N'$.  The solid lines going up and down in
Fig.~\ref{fig:1ph-pi} follow the standard many body nomenclature and stand for
particle and hole states respectively.
The $W-$self-energy corresponding to this diagram (actually 49
diagrams) is readily evaluated and gives\footnote{In
Eq.~(\ref{eq:W1pibis}), it is necessary to subtract the free space
contribution, i.e., the one that survives for vanishing nuclear
densities. This contribution will renormalize free space couplings and
masses. To obtain Eq.~(\ref{eq:W1pi}), we have neglected the
contribution of the antiparticle pole ($p^0=- E(\vec{p}\,)-{\rm
i}\epsilon$) in the $p^0$ integration, this also gets rid of the vacuum
contribution that needed to be subtracted.} (we will label it as 1p1h$1\pi$)
\begin{eqnarray}
-{\rm i} \Pi_{W;1{\rm ph}1\pi}^{\mu\nu}(q)&=&  -{\rm i} \left
(\frac{g}{2\sqrt 2} \right )^2\sum_{N,N',\lambda} \int \frac{ d^4
  k_\pi}{(2\pi)^4} \int\frac{d^4p}{(2\pi)^4} G(p\,;\rho_N)
G(p'\,;\rho_{N'}) D_\pi(k_\pi)  \times \nonumber \\
&\times& {\rm
  Tr}\left((\slashchar{p}+M)\gamma^0j^{ \mu
  \dagger}_A\gamma^0(\slashchar{p}'+M)j^\nu_A
\right) \label{eq:W1pibis} \\
&=&\left (\frac{g}{2\sqrt 2} \right )^2\sum_{N,N',\lambda} \int \frac{
d^4 k_\pi}{(2\pi)^4}  \int\frac{d^3p}{(2\pi)^3}\frac{1}{2E(\vec{p}\,)}
\frac{1}{2E(\vec{p}+\vec{q}-\vec{k}_\pi)}\frac{n_N(\vec{p}\,)[1-n_{N^\prime} 
(\vec{p}+\vec{q}-\vec{k}_\pi)]}{q^{0}-k_\pi^{0}+E(\vec{p}\,)-E(\vec{p}\,')+i\epsilon}
 \times \nonumber \\
&  \times & D_\pi(k_\pi) {\rm Tr}\left((\slashchar{p}+M)\gamma^0j^{ \mu
   \dagger}_A\gamma^0(\slashchar{p}'+M)j^\nu_A \right)  + \left[ (q-k_\pi)\leftrightarrow -(q-k_\pi) \right ] \label{eq:W1pi}
\end{eqnarray}
\noindent
where $p'=p+q-k_\pi$, $j^\mu_A$ is the amputated
  amplitude\footnote{The dependence of $j^\mu_A$ on the $N,N',
  \lambda$ channel is understood and it is not made explicit.} for the
  $W^+N\rightarrow N' \pi^\lambda$ process, which is obtained by summing
  up the contributions of all diagrams of the right panel of
  Fig.~(\ref{fig:diagramas}). The contribution to $j^\mu_A$ of each
  diagram is given by their relation to the full amplitudes given in
  Eq.~(51) of Ref.~\cite{Hernandez:2007qq},
\begin{equation}
j^\mu_{{\rm cc}+}\Big|_i = 
\bar u(\vec{p}\,') j^\mu_{A_i} (p,q,p'=p+q-k_\pi,k_\pi) 
u (\vec{p}\,),\quad i = \Delta P, C\Delta P, NP, CNP, CT, PP,
PF. \label{eq:defjA} 
\end{equation}
The indices $N, N'$ in Eq.~(\ref{eq:W1pi}) stand for the hole and
particle nucleon states respectively and $n_N (\vec{p}\,) =
\Theta(k_F^N-|\vec{p}\,|)$ is the occupation number in the Fermi local
sea, with $k_F^N(r)= (3\pi^2\rho_N(r))^{1/3}$ and $\rho_N(r)$ the
density of nucleons of a particular species $N=n$ or $p$ ($\rho(r) =
\rho_p(r) + \rho_n(r) $), normalized to the number of protons or
neutrons. Besides, $E (\vec{p})$ is the energy of the nucleon
$\sqrt{\vec{p}\,^2 + M^2}-k_F^2/2M$, with $M$ its mass and $k_F(r)=
(3\pi^2\rho(r)/2)^{1/3}$, and $D_\pi$ is the pion propagator
\begin{equation}
 D_{\pi}(k_\pi)=\frac{\displaystyle{1}}
  {\displaystyle{k_\pi^{2}-m_{\pi}^{2}+i\epsilon}}
\end{equation}
with $m_\pi$ the mass of the pion. Besides, the nucleon propagator 
in the medium reads,
\begin{eqnarray}
S(p\,; \rho) &=& (\slashchar{p}+M) G(p\,; \rho) \\
G(p\,; \rho) &=& \frac{1}{p^2-M^2+{\rm
    i}\epsilon}  + {\rm
  i}\frac{\pi}{E(\vec{p}\,)}n(\vec{p}\,)\delta(p^0-E(\vec{p}\,))
\label{eq:Gpbis}\\
&=& \frac{1}{p^0+ E(\vec{p}\,) + {\rm
    i}\epsilon} \left ( \frac{n(\vec{p}\,)}{p^0- E(\vec{p}\,) - {\rm
    i}\epsilon} + \frac{1-n(\vec{p}\,)}{p^0- E(\vec{p}\,) + {\rm
    i}\epsilon}  \right )  
  \label{eq:Gp}
\end{eqnarray}

A further simplification can be done by evaluating the $j^\mu_A$
 amplitudes at an average momentum, which allows to take the
 spin trace in Eq.~(\ref{eq:W1pi}) out of the $d^3 \vec{p}$
 integration. This latter integration can be now done, and it gives,
 up to some constants, the Lindhard function,
 $\overline {U}_R(q-k_\pi,k_F^N,k_F^{N'})$ (see appendix B of
 Ref.~\cite{Nieves:2004wx}).  We take $\langle |\vec{p}\,|
 \rangle= \sqrt{\frac{3}{5}} k_F^N$ and a direction orthogonal to the
 plane defined by the pion and the virtual gauge boson. Within this
 approximation, we find
\begin{eqnarray}
-{\rm i} \Pi_{W;1{\rm ph}1\pi}^{\mu\nu}(q)&=& \left (\frac{g}{2\sqrt 2} \right
)^2\frac{1}{4M^2}\sum_{N,N',\lambda} \int \frac{ d^4 k_\pi}{(2\pi)^4}
 D_\pi(k_\pi)\overline {U}_R(q-k_\pi,k_F^N,k_F^{N'}) 
A^{\mu\nu}\left[\langle p\rangle,q,p'=\langle p\rangle+q-k_\pi,k_\pi\right]
 \\ 
A^{\mu\nu} &=& \frac12 {\rm
Tr}\left(\left(\langle\slashchar{p}\rangle+M\right)\gamma^0\langle j^{ \mu
\dagger}_A\rangle \gamma^0\left(\langle\slashchar{p}\rangle+\slashchar{q}
-\slashchar{k}_\pi+M\right)\langle j^\nu_A\rangle \right) \label{eq:1ph1pi}
\end{eqnarray}
\noindent
where $\langle j^\nu_A\rangle$ stands for $j^\nu_A$ calculated with
the average hole momentum $\langle \vec{p}\,
 \rangle$.  To find the contribution to the hadron tensor
 $W^{\mu\sigma}$ of the many body diagrams depicted in
 Fig.~\ref{fig:1ph-pi}, we remind that  by construction 
\begin{equation}
A^{\mu\nu} = A^{\mu\nu}_s + {\rm i} A_a^{\mu\nu}\label{eq:sym}
\end{equation}
with $A^{\mu\sigma}_s$ ($A^{\mu\sigma}_a$) real symmetric
(antisymmetric) tensors, and thus 
\begin{equation}
{\rm Im}\left [ \Pi_{W;1{\rm ph}1\pi}^{\mu\nu} 
+ \Pi_{W;1{\rm ph}1\pi}^{\nu\mu} \right ] = 2  {\rm Im} \Pi_{W;1{\rm
    ph}1\pi}^{\mu\nu}\Big|_s, \qquad   {\rm Re}\left [ \Pi_{W;1{\rm ph}1\pi}^{\mu\nu} 
- \Pi_{W;1{\rm ph}1\pi}^{\nu\mu} \right ] = -2  {\rm Im} \Pi_{W;1{\rm
    ph}1\pi}^{\mu\nu}\Big|_a
\end{equation}
where $\Pi_{W;1{\rm ph}1\pi}^{\mu\nu}\Big|_{s(a)}$ is defined as in
Eq.(\ref{eq:1ph1pi}), but replacing the full tensor $A^{\mu\nu}$ by
its symmetric (antisymmetric) $A^{\mu\nu}_{s(a)}$ parts.  The imaginary
part of $\Pi_{W;1{\rm ph}1\pi}^{\mu\nu}\Big|_{s(a)}$ can be obtained
by following the prescription of the Cutkosky's rules. In this case we
cut with a straight horizontal line the intermediate particle and hole
states and the pion. Those states
are then placed on shell by taking the imaginary part of the
propagator.  Technically the rules to obtain ${\rm Im}
\Pi_{W;1{\rm ph}1\pi}^{\mu\nu}\Big|_{s(a)}$ reduce 
to making the substitutions:
\begin{eqnarray}
\Pi_W^{\mu\nu}(q)&\rightarrow& 2{\rm i}{\rm Im}\Pi^{\mu\nu}_W(q)\Theta(q^0)\\
D_\pi (k_\pi) &\rightarrow& 2{\rm i} {\rm Im} D_\pi (k_\pi)
\Theta(k_\pi^0)  = -2\pi
{\rm i} \delta(k_\pi^2-m_\pi^2) \Theta(k_\pi^0) \\
\overline {U}_R(q-k_\pi,k_F^N,k_F^{N'}) &\rightarrow& 2{\rm i}{\rm
  Im}\overline {U}_R(q-k_\pi,k_F^N,k_F^{N'}) \Theta(q^0-k_\pi^0) \
\end{eqnarray}           
Thus, we readily obtain
\begin{equation}
W^{\mu\nu}_{1{\rm ph}1\pi}(q) = -\Theta(q^0) \frac{1}{2M^2} \int
\frac{d^3r }{2\pi} 
\sum_{N,N',\lambda} \frac{d^3k_\pi}{(2\pi)^3}
\frac{\Theta(q^0-k_\pi^0)}{2\omega(\vec{k_\pi}\,)} {\rm
  Im}\overline {U}_R(q-k_\pi,k_F^N,k_F^{N'}) A^{\nu\mu} \label{eq:1p1hpi-def}
\end{equation}
with $\omega(\vec{k_\pi}\,)$ the pion on-shell energy.  The
approximation done saves a considerable amount of computational time
since there are analytical expressions for ${\rm Im}\overline
{U}_R(q-k_\pi,k_F^N,k_F^{N'}) $~\cite{Nieves:2004wx}\footnote{In the small
density limit $ {\rm Im}\overline {U}_R(q,k_F^N,k_F^{N'}) \simeq - \pi
\rho_N M\delta \left(q^0+M -
\sqrt{M^2+\vec{q\,}^2}\right)/\sqrt{M^2+\vec{q\,}^2}$. Substituting
this into Eq.~(\ref{eq:1p1hpi-def}) one obtains
\begin{eqnarray}
\lim_{\rho\to 0}W^{\mu\nu}_{1{\rm ph}1\pi} \sim Z W^{\mu\nu}_{W^+p \to
p\pi^+} + N \left (W^{\mu\nu}_{W^+n \to
p\pi^0} +W^{\mu\nu}_{W^+n \to
n\pi^+} \right ) 
\end{eqnarray}
being $Z$ and $N$ the number of protons and neutrons of the nucleus,
and $W^{\mu\nu}_{W^+N \to N'\pi^\lambda}$ the hadronic tensor for CC
pion production  off the nucleon (see Eq.~(4) of
Ref.~\cite{Hernandez:2007qq}). In this way, the strict impulse
approximation is recovered. By performing the integral in
Eq.~(\ref{eq:1p1hpi-def}), one accounts for Pauli blocking and 
for Fermi motion.}.

\subsection{The dip region: 2p2h absorption}

In the two previous subsections we have discussed the dominant
contributions to the inclusive $\nu$,$\bar{\nu}$ nucleus CC cross
section at low energies, namely QE scattering and pion production.  In
this subsection, we present a model for 2p2h mechanisms, which could be
very relevant in the description of the region of transferred energies
above the quasielastic and below the $\Delta-$resonance peaks (the dip
region).
 
\subsubsection{2p2h mechanisms driven by the longitudinal part of the
 effective spin--isospin ph--ph interaction}

\begin{figure}
\includegraphics[height=.16\textheight]{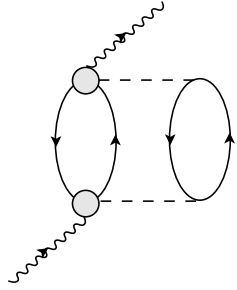}
 \caption {$W-$self-energy obtained from the one in 
 Fig.~\ref{fig:1ph-pi} when the pion line is allowed to excite a
 particle--hole.}
\label{fig:2ph}
\end{figure}

Let us consider again to the generic diagram of weak pion production  of 
Fig.~\ref{fig:1ph-pi} and allow 
the pion to excite a particle--hole. This leads us to the diagram of
Fig.~\ref{fig:2ph}.  
This is still a generic diagram which actually contains 49 diagrams
when in the shaded circle we put each one of the terms of the $W N
\rightarrow \pi N$ amplitude of Fig.~\ref{fig:diagramas}. The diagrams
in Fig.~\ref{fig:2ph} contribute to ${\rm Im} \Pi^{\mu\nu}_W$
according to Cutkosky rules when they are cut by a horizontal line,
and the 2p2h are placed on shell. The contribution of the diagram of
Fig.~\ref{fig:2ph} can be obtained from that of Fig.~\ref{fig:1ph-pi},
given in Eq.~(\ref{eq:W1pi}), by replacing ($\Lambda_\pi=1.2~{\rm
GeV},~f^2_{\pi NN}/4\pi=0.08$)
\begin{equation}
D_\pi (k_\pi) \to D_\pi^2 (k_\pi) F_\pi^4(k_\pi) \frac{f^2_{\pi NN}}{m_\pi^2}
\vec{k}_\pi^2 U_\lambda (k_\pi), ~~~F_\pi(k)=
\frac{\Lambda_\pi^2-m_\pi^2}{\Lambda_\pi^2-k_\pi^2
},
\end{equation}
where $U_\lambda$ is the Lindhard function for a particle--hole
excitation by an object of charge $\lambda$: this is, twice
$\bar{U}^{p,n}_R$ or $\bar{U}^{p,n}_R$ for the excitation by a charged
pion or $\bar{U}^{p,p}_R+\bar{U}^{n,n}_R$ for the excitation by a
neutral pion.  The pion form factor $F^4_\pi (k_\pi)$ appears because
now the pions are off shell.

We can again simplify the expression by taking an average nucleon
momentum of the Fermi sea to evaluate the amputated amplitudes for the
$W^+N\rightarrow N' \pi^\lambda$ process. This allows us to factorize
the Lindhard function and following the prescription of the Cutkosky's
rules we get, 
\begin{eqnarray}
W_{{\rm 2p2h}}^{\mu\nu}(q)&=& \Theta(q^0) \frac{1}{M^2}\int 
\frac{d^3r}{2\pi}\sum_{N,N',\lambda} \int \frac{ d^4 k_\pi}{(2\pi)^4}
\Theta(q^0-k_\pi^0) F_\pi^2(k_\pi) 
{\rm Im} \overline {U}_R(q-k_\pi,k_F^N,k_F^{N'}) A^{\nu\mu}  \times \nonumber
 \\
&&\times  D^2_\pi(k_\pi) F_\pi^2(k_\pi) \frac{f^2_{\pi NN}}{m_\pi^2}
\vec{k}_\pi^2 \Theta(k_\pi^0) {\rm Im} U_\lambda (k_\pi)\label{eq:W2p2h}
\end{eqnarray}
Next, we have implemented several improvements that account for well
established many body corrections: 
\begin{enumerate}
\item In the above expression of Eq.~(\ref{eq:W2p2h}), we have replaced 
\begin{equation}
 D^2_\pi(k_\pi)F_\pi^2(k_\pi) \frac{f^2_{\pi NN}}{m_\pi^2}
\vec{k}_\pi^2 \, {\rm Im} U_\lambda (k_\pi) \to {\rm Im}
\left ( \frac{1}{k_\pi^2-m_\pi^2-\Pi (k_\pi) } \right )  = \frac{{\rm Im}\Pi}{|k_\pi^2-m_\pi^2-\Pi (k_\pi)|^2}
\label{eq:dy-2p2h}
\end{equation}
where for the selfenergy of a pion of charge $\lambda$, we have
taken~\cite{Nieves:1991ye} 
\begin{equation}
\Pi (k_\pi) = F_\pi^2(k_\pi) \frac{f^2_{\pi NN}}{m_\pi^2}
\vec{k}_\pi^2  \frac{ U(k_\pi)}{1-\frac{f^2_{\pi
      NN}}{m_\pi^2} g'  U(k_\pi)} \label{eq:uu's}
\end{equation}
where 
\begin{equation}
U (k_\pi) = U_N (k_\pi) + U_\Delta (k_\pi) \label{eq:def_lin}
\end{equation}
is the non-relativistic Lindhard function for ${\rm ph}\,+\,\Delta$h
excitations\footnote{The functions $U_N$ and $U_\Delta$ are defined,
e.g., in Eqs. (2.9) and (3.4) of
Ref.~\protect\cite{GarciaRecio:1987ik}. $U_N$ incorporates a factor
two of isospin with respect to $\bar U_R$, such that ${\rm Im} U_N = 2
{\rm Im} \bar U_R$ for symmetric nuclear matter, up to relativistic
corrections.} including direct and
crossed bubbles \cite{Oset:1987re, GarciaRecio:1987ik}, in contrast to
$\bar U_R$ which only contains the direct bubble of a particle--hole
excitation (the only one which contributes to ${\rm Im} U_N $ for $q^0
> 0$). When evaluating ${\rm Im}\,\Pi(k_\pi)$ in the numerator of
Eq.~(\ref{eq:dy-2p2h}) we have not considered the
 part that arises from putting the $\Delta$h excitation
on-shell that would correspond to a 2p2h+$1\pi$ mechanism. We expect
this latter contribution to be small at the considered energies.  Note
also, that by using $U$ to compute the pion selfenergy,  we have
neglected small relativistic and $\rho_p\ne \rho_n$ corrections.
By means of Eq.~(\ref{eq:dy-2p2h}), we have implemented the Dyson
re-summation of the pion selfenergy, and have improved on this latter
one by incorporating the Lorentz-Lorenz effect, driven by the short
range Landau Migdal parameter $g'$~\cite{Nieves:1993ev}, going in
this way beyond 1p1h excitation\footnote{It corresponds to replace the
ph excitation of the right-hand in Fig.~\ref{fig:2ph} by a series of
RPA excitations through ph and $\Delta$h excitations, driven by the
longitudinal part of the effective spin-isospin interaction. In
Subsect.~\ref{sec:2p2h-transverse}, we do something similar for the
case of 2p2h mechanisms driven by $\rho-$meson exchange, and there we
show graphically in Fig.~\ref{fig:rpa-t} the RPA series, in that case
induced by the transverse part of the effective spin-isospin
interaction.}  in the evaluation of $\Pi(k_\pi)$. We have used $g'=0.63$, as in
previous works~\cite{Nieves:1993ev,Nieves:1991ye,Nieves:2004wx, Gil:1997bm}.

\item 
  Let us now pay attention to the diagram of Fig.~\ref{fig:PNterm},
  which is already implicit in the generic diagram of Fig.~\ref{fig:2ph}
  when the $NP$ amputated amplitude is considered in both weak
  vertices.  This $W-$selfenergy contribution can be obtained from the QE
  1p1h excitation term (first of the diagrams depicted in
  Fig.~\ref{fig:fig1}) by dressing up the nucleon propagator of the
  particle state in the ph excitation. Indeed, this, among other
  contributions, was already taken into account in the QE study
  carried out in our previous work of Ref.~\cite{Nieves:2004wx}, since
  there dressed nucleon propagators deduced from a  realistic nucleon
  selfenergy~\cite{FernandezdeCordoba:1991wf}  were used.
\begin{figure}
\includegraphics[height=.16\textheight]{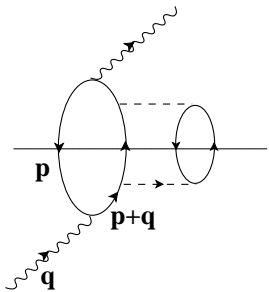}
 \caption {$W-$selfenergy diagram obtained from the QE 1p1h excitation
 term (first of the diagrams depicted in Fig.~\ref{fig:fig1}) by
 dressing up the nucleon propagator of the particle state in the ph
 excitation.}
\label{fig:PNterm}
\end{figure}
To avoid double counting, we subtract
the contribution of the $NP$--$NP$ diagram of
Fig.~\ref{fig:PNterm} from Eq.~(\ref{eq:W2p2h}). 

\item When in one of the weak vertices of Fig.~\ref{fig:2ph}, the $NP$
term is considered, the prescription of taking an average nucleon
momentum of the Fermi sea used to obtain Eq.~(\ref{eq:W2p2h}) turns
out to be not appropriated. The reason is that when placing the 2p2h
excitation on shell, through Cutkosky rules, we still have the nucleon
propagator with momentum $p + q$ (this is part of the amputated
amplitude $j^\mu_A\Big|_{NP}$). This propagator can be still placed on shell
for a virtual $W$ and thus, there exists a single pole in the $d^3p$
integration\footnote{This cut will also contribute to the nuclear
response to the weak probe. But, while it will affect to the QE
region, it is  expected to be small and
considerably difficult to calculate from the computational point of
view (see Eq.~(80) of Ref.~\cite{Gil:1997bm}). Thus, for the sake of
simplicity we have not considered such contribution.}. In this
situation, one can not take an average for $\vec{p}$, as we have
implicitly assumed in Eq.~(\ref{eq:W2p2h}), and we have improved such
prescription as follows. In this latter equation, it appeared the
tensor $A^{\nu\mu}$, which in turn is defined in Eq.~(\ref{eq:1ph1pi})
by using an average for the hole three momentum $\vec{p}$ to calculate
both, the amputated $WN \to N\pi$ amplitudes $j_A^\mu$ and
$\slashchar{p}$, that also appears in the trace that defines
$A^{\nu\mu}$. Instead of this, we have computed an average of the
whole trace. To this end, we have numerically performed the integral
over the angle formed by $\vec{p}$ and $\vec{q}$, using still an
average for the modulus of $\vec{p}$ and taking this momentum in the
XZ plane (recall that $\vec{q}$ defines the Z-axis). All pathologies
arise from the $p+q$ nucleon propagator hidden in the amputated
amplitudes, which can be put on the mass shell, and thus the
contribution of these diagrams depends critically on the angle formed
by $\vec{p}$ and $\vec{q}$, while it shows a very smooth dependence on
the rest of kinematical variables of the hole momentum
$\vec{p}$. Thanks to the approximations of using an average for the
modulus of $\vec{p}$ and fixing the $(\vec{p}, \vec{q}\,)-$plane , we
avoid to perform two nested integrals, with the obvious benefit in
computation time. We have checked that the results are accurate
at the level of 5--10\%. To be more specific, in Eq.~(\ref{eq:W2p2h}),
we have replaced $A^{\nu\mu}$ by
\begin{equation}
A^{\nu\mu} \Rightarrow  \frac12 \int^{+1}_{-1} d\mu \frac12 {\rm
Tr}\left(\left(\slashchar{p}+M\right)\gamma^0 j^{ \nu
\dagger}_A \gamma^0\left(\slashchar{p}+\slashchar{q}
-\slashchar{k}_\pi+M\right) j^\mu_A \right) \label{eq:avg}
\end{equation}
with $\mu = \vec{q}\cdot\vec{p}/|\vec{q}\,||\vec{p}\,|$. To speed up
the numerical integration, we have also given a small width ($\sim 10$ MeV)
to the $p+q$ nucleon. Results do not depend significantly on this
choice.  

For consistency, we have also performed this angular average for all
contributions implicit in Fig.~\ref{fig:2ph}, though the prescription
of using an average for $\vec{p}$ leads to accurate results in all
cases except those involving the $NP$ amputated amplitude 
discussed above.

\item 
In the terms involving the $NP$, amputated amplitude
(interferences with the rest of amplitudes of Fig.~\ref{fig:fig1}),
there always appears a pion emitted after the $WN$ vertex that couples
to the second ph excitation (see for instance the line labeled as
$\pi$ in Fig.~\ref{fig:PNinter-term}). There, one is assuming a pion
exchange interaction among the two ph excitations. We have improved on
that, and have replaced it by an effective longitudinal interaction,
$V_l$, 
\begin{eqnarray}
V_l(k) &=& \frac{f^2_{\pi NN}}{m^2_\pi}\left
\{ F^2_\pi(k)
\frac{\vec{k}^2}{k^2-m_\pi^2} + g^\prime_l(k)\right \},\label{eq:st2x0}
\end{eqnarray}
which besides pion exchange includes SRC driven by the Landau
Migdal parameter $g^\prime_l(k)$ (see 
Refs.~\cite{Oset:1987re,Nieves:1993ev,Nieves:1991ye}). To achieve this, we
have multiplied the amputated amplitude $j^\mu_{A_{NP}}$ by a suitable factor,
\begin{equation}
j^\mu_{A_{NP}} \Rightarrow j^\mu_{A_{NP}} \times 
\left ( 1+ \frac{g'_l}{F^2_\pi D_\pi \vec{k}_\pi^2}\right) 
\end{equation}
We have taken
the same prescription also for those terms that include the $CNP$,
$\Delta P$ and $C \Delta P$ amputated amplitudes. 

We have also considered the transverse channel interaction, $V_t$,
\begin{eqnarray}
V_t(k) &=& \frac{f^2_{\pi NN}}{m^2_\pi}\left
\{   C_\rho F^2_\rho(k)
\frac{\vec{k}^2}{k^2-m_\rho^2} + g^\prime_t(k)\right \}, \quad
C_\rho=2,~~F_\rho(k)=\frac{\Lambda_\rho^2-m_\rho^2}{\Lambda_\rho^2-k^2 },~~\Lambda_\rho=2.5~{\rm GeV} \label{eq:st2x3}
\end{eqnarray}
of the effective spin-isospin interaction among
the two ph excitations. Here, $m_\rho=0.77~{\rm GeV}$.
The SRC functions $g^\prime_l$ and
$g^\prime_t$ have a smooth
$k-$dependence~\cite{Oset:1981ih,Oset:1987re}, which we will not
consider here\footnote{This is justified because taking into account
the $k-$dependence leads to minor changes for low and intermediate
energies and momenta, where this effective ph-ph interaction should be
used.}, and thus we will take
$g^\prime_l(k)=g^\prime_t(k)=g^\prime=0.63$, as it was done in
the study of inclusive nuclear electron scattering carried out in
Ref.~\cite{Gil:1997bm}, and also in the previous work on the QE region
of Ref.~\cite{Nieves:2004wx}.
 To account for such contribution to the 2p2h
absorption cross section  is slightly more complicated, because
the tensor $A^{\mu\nu}$ does not account for $\rho-$meson production
in the primarily weak vertex.  Details will be given in
Subsect.~\ref{sec:2p2h-transverse}.
\begin{figure}[t]
\includegraphics[height=.16\textheight]{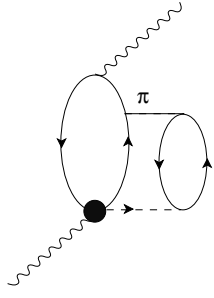}
 \caption {$W-$selfenergy diagrams in which one of the vertices
 contains the $NP$ term of the $WN \to N \pi$ amplitude, while the
 other one (filled circle) contains all terms except that one.}
\label{fig:PNinter-term}
\end{figure}
\end{enumerate}

\begin{figure}[htb]
\includegraphics[height=.16\textheight]{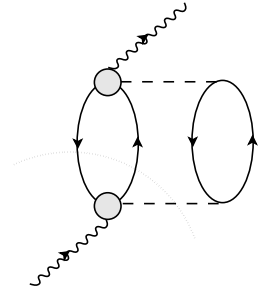}
 \caption {Same as  Fig.~\ref{fig:2ph}, showing the cut which places 
    one particle--hole and the pion on shell.}
\label{fig:2phCQ}
\end{figure}

The cut which places the two ph on shell in the diagrams of
Fig.~\ref{fig:2ph} is not the only possible one. In
Fig.~\ref{fig:2phCQ}, we show a different cut (dotted line) which
places one ph and the pion on shell.
 As done for
real~\cite{Carrasco:1989vq} and virtual~\cite{Gil:1997bm} photons, we
neglect this contribution in the non resonant terms, because at low energies
where these pieces are important, the $(W, \pi)$
channel is small and at high energies where the $(W, \pi)$
contribution is important, this channel is dominated by the $\Delta$
excitation and there this correction will be properly  incorporated.

\begin{figure}[htb]
\includegraphics[height=.16\textheight]{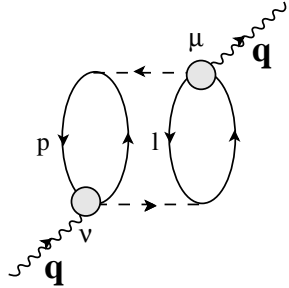}
 \caption {Two particle--two hole $W-$selfenergy Feynman diagram 
 where the outgoing gauge boson couples to the second nucleon. }
\label{fig:2ph2bubbles}
\end{figure}

We have also considered two body diagrams, where each $W$ boson
couples to different bubbles (Fig.~\ref{fig:2ph2bubbles}). Its
contribution to the hadron tensor, taking  average momenta for
both hole nucleon momenta in first place, reads
\begin{eqnarray}
W_{{\rm 2p2h}-{2\rm b}}^{\mu\nu}(q)&=& -\Theta(q^0) \frac{1}{2 \sqrt 2
M^4}\frac{f^2_{\pi NN}}{m_\pi^2} \int \frac{d^3r}{2\pi} \int \frac{
d^4 k_\pi}{(2\pi)^4} \Theta(k_\pi^0) \Theta(q^0-k_\pi^0)
F_\pi^4(k_\pi) \times \nonumber \\ && \times D_\pi (k_\pi) D_\pi
(k_\pi-q) {\rm Im}\, \overline {U}_R(k_\pi)\, {\rm Im}\, \overline
{U}_R(q-k_\pi) A^{\nu\mu}_ {2\rm b} \\
A^{\mu\nu}_ {2\rm b} &=& \frac12 {\rm
Tr}\left[\left(\langle\slashchar{p}\rangle+M\right)
\left(\slashchar{k}_\pi-\slashchar{q}\right) \gamma_5 \left(\langle\slashchar{p}\rangle+\slashchar{q}
-\slashchar{k}_\pi+M\right) \underbrace{\langle j^\nu_A\rangle}_{W^+p\to p
  \pi^+} \right] \times \nonumber \\
&\times&  \frac12 {\rm
Tr}\left[\left(\langle\slashchar{l}\rangle+M\right)\gamma^0\underbrace{\langle j^{ \mu
\dagger}_A\rangle}_{W^+n\to p
  \pi^0}  \gamma^0 \left(\langle\slashchar{l}\rangle+\slashchar{k}_\pi+M\right)
\slashchar{k}_\pi \gamma_5\right] + \left( W^+n\to p \pi^0\right) \left(
W^+p\to p \pi^+\right) \nonumber  \\
&-&  
\left( W^+n\to p \pi^0\right) \left( W^+n\to n \pi^+\right)- \left(
W^+n\to n \pi^+\right) \left( W^+n\to p \pi^0\right)  \label{eq:W2p2h2bubbles}  
\end{eqnarray}
To compute this moderately small term, we have taken a proton--neutron
 symmetric Fermi sea, i.e., $k_F^p = k_F^n$.  We have labeled by $p$
 and $l$, the left and right bubble hole four momenta, and to compute
 $\langle j^\nu_A\rangle$ and $\langle j^{ \mu \dagger}_A\rangle$, the
 pion momenta are $k_\pi$ and $k_\pi-q$, respectively. Besides, we
 write explicitly in Eq.~ (\ref{eq:W2p2h2bubbles}) the sum over all
 charge combinations.

Finally,  we have improved on our evaluation of
$W_{{\rm 2p2h}-{2\rm b}}^{\mu\nu}$ given in
Eq.~(\ref{eq:W2p2h2bubbles}) by implementing  also here the
refinements 3 and 4, previously discussed, devoted to the
improvements in the computation of $W_{{\rm 2p2h}}$. However, as in
Ref.~\cite{Gil:1997bm} for the inclusive electron-nucleus case,
we have only considered the contributions stemming from the
longitudinal  part of the effective spin--isospin ph-ph interaction, driven
by pion exchange (depicted in Fig.~\ref{fig:2ph2bubbles}), and we have
  neglected those induced by the transverse one.

\subsubsection{2p2h mechanisms driven by the transverse part of the
 effective spin--isospin ph--ph interaction}
\label{sec:2p2h-transverse}

In this subsection we evaluate the contribution to the
weak nuclear response of the 2p2h absorption terms driven by the transverse
part of the effective  spin--isospin ph--ph potential used in previous
studies on electron~\cite{Gil:1997bm}, photon~\cite{Carrasco:1989vq}
and pion~\cite{Oset:1981ih,Salcedo:1987md, Nieves:1993ev,
Nieves:1991ye, Albertus:2001pb} interaction with nuclei. In the model
of Ref.~\cite{Oset:1981ih}, this transverse interaction arises from
$\rho-$exchange modulated by SRC. The major difficulty here, as
compared with the previous works mentioned above, arises from the fact
that we are using a relativistic description of the weak transition
process. Thus, the first step is to model $NN\rho$ and $N\Delta\rho$
relativistic Lagrangians, which give rise, in the non-relativistic
limit, to the transverse potential of Eq.~(\ref{eq:st2x3}). A convenient
set of interaction Lagrangians is,
\begin{eqnarray}
{\cal L}_{NN\rho} &=& \frac{f_{\pi NN}}{m_\pi}\sqrt{C_\rho} \bar\Psi
\sigma_{\mu\nu} \partial^\mu \vec{\rho}^{\,\nu
} \cdot \vec{\tau}\, \Psi \nonumber \\
{\cal L}_{N\Delta\rho} &=& -{\rm i} \frac{f^*_{\pi N\Delta}}{m_\pi} 
\sqrt{C_\rho} \bar\Psi_\nu \gamma_5 \gamma_\mu \vec{T}^\dagger\Psi \left (\partial^\mu \vec{\rho}^{\,\nu} - \partial^\nu \vec{\rho}^{\,\mu}\right) + {\rm h.c.}
\end{eqnarray}
where $\Psi= \left (\begin{array}{c}p\cr n\end{array}\right )$ is the
nucleon field, $\vec{\rho}^{\,\nu}$ is the $\rho-$meson Proca
field\footnote{Here 
$\rho^{\,\mu}=(\rho^\mu_1-{\rm i}\, \rho^\mu_2)/\sqrt{2}$ creates a
$\rho^-$ from the vacuum or annihilates a $\rho^+$ and the
$\rho_3^\mu$ field creates or annihilates a $\rho^0$.}, $\Psi_\nu$ is
a Rarita Schwinger $J^\pi = 3/2^+$ field, $\vec{T}^\dagger$ is the
isospin transition operator\footnote{It is a vector under isospin
rotations and its Wigner-Eckart irreducible matrix element is taken to
be one.}, $\vec\tau$ are the isospin Pauli matrices and
$f^*_{\pi N\Delta}=2.14$. 

Next, we consider the $W^+N \to N' \rho$ process and find the $NP$,
$CNP$, $\Delta P$ and $C\Delta P$ amputated amplitudes obtained from
the above Lagrangians. The Feynman diagrams for these four amplitudes
are like those depicted in the right panel of Fig.~\ref{fig:diagramas}
by replacing the outgoing pion by a $\rho-$meson. We will denote the
amputated amplitudes by $t^{\mu\alpha}_{A_i}$. They are defined by
\begin{equation}
t^\mu_{cc+}\Big|_i = \bar u(\vec{p}\,') t^{\mu\alpha}_{A_i} (p,q,p'=p+q-k_\rho,k_\rho) 
u (\vec{p}\,)\epsilon^*_\alpha (k_\rho),\quad i = \Delta P, C\Delta P, NP, CNP
\end{equation}
with $\epsilon^*_\alpha$, the $\rho-$meson polarization vector and
$t^\mu_{cc+}\Big|_i$ the full $W^+N \to N' \rho$ amplitude for each
mechanism. One readily finds that $t^\mu_{cc+}\Big|_i$ can be obtained
from the $W^+N \to N' \pi$ amplitudes $j^\mu_{cc+}\Big|_i$, given in Eq.~(51)
of Ref.~\cite{Hernandez:2007qq}, with the following replacements
deduced  from the appropriate meson-$NN$ and meson-$N\Delta$ vertices,
\begin{eqnarray}
NP~{\rm and}~ CNP~{\rm terms:}\qquad 
\slashchar{k}_\pi \gamma_5 &\Leftrightarrow&
\sqrt{C_\rho} k^\eta\sigma_{\eta \sigma }  \epsilon^{*\sigma}
\nonumber \\
\Delta P~{\rm term:}\qquad k_\pi^\alpha P_{\alpha\beta} \#^\beta
&\Leftrightarrow& {\rm i} \sqrt{C_\rho} \gamma_5 \left( \slashchar{k}
P_{\sigma \beta }- k^\alpha \gamma_\sigma P_{\alpha\beta}  \right)\#^\beta
  \epsilon^{*\sigma} 
\nonumber \\
C \Delta P~{\rm term:}\qquad  \#^\alpha k_\pi^\beta P_{\alpha\beta}
&\Leftrightarrow& -{\rm i} \sqrt{C_\rho} \#^\alpha \left( P_{\alpha \sigma}
\gamma_5  \slashchar{k}
- P_{\alpha\eta} k^\eta \gamma_5 \gamma_\sigma
\right)\epsilon^{*\sigma}  
\end{eqnarray}
where $k$ is now the $\rho-$meson momentum ($p+q=p'+k$).
\begin{figure}
\includegraphics[height=.16\textheight]{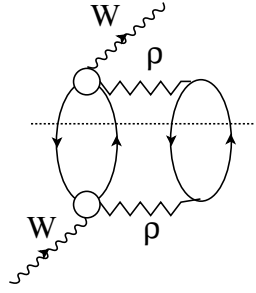}
 \caption {Two particle--two hole $W-$selfenergy Feynman diagram
 driven by $\rho-$exchange. The cut (dotted line) that places the 2p2h
 on-shell is also displayed. The empty circle contains the direct and
 crossed nucleon and $\Delta-$pole terms of the $WN \to N\rho$
 amplitude.}
\label{fig:rho-vt}
\end{figure}
Now, we can  compute the contribution of the 16 diagrams of
Fig.~\ref{fig:rho-vt} to the inclusive neutrino-nucleus cross section,
when the two ph excitations are put on shell. The corresponding
contribution to the hadron tensor reads, with the approximations
discussed in the previous sections in order to factorize 
two Lindhard functions,  
\begin{eqnarray}
W_{{\rm 2p2h}-t}^{\mu\nu}(q)&=& -\Theta(q^0) \frac{1}{M^2}\int 
\frac{d^3r}{2\pi}\sum_{N,N',\lambda} \int \frac{ d^4 k}{(2\pi)^4}
\Theta(q^0-k^0) F_\rho^2(k) 
{\rm Im} \overline {U}_R(q-k,k_F^N,k_F^{N'}) B^{\nu\mu}  \times \nonumber
 \\
&&\times  D^2_\rho(k) F_\rho^2(k)  C_\rho^2 \frac{f^2_{\pi NN}}{m_\pi^2}
\vec{k}_\pi^2 \Theta(k^0) {\rm Im} U_\lambda (k)\label{eq:W2p2h-t}
\end{eqnarray}
with\footnote{ This is part of the $\rho-$meson
propagator, that reads $(-g^{\mu\nu}+k^\mu k^\nu/m^2_\rho ) D_\rho$.
Actually,  only the piece proportional to $g^{\mu\nu}$ contributes
since $t^{\mu\alpha}_{A_i} k_\alpha =0$.} $D_\rho (k) =
1/(k^2-m^2_\rho)$.   The  form
factor $F^4_\rho (k)$, Eq.~(\ref{eq:st2x3}), appears because the
$\rho's$ are off shell. Also here, when placing the
2p2h excitations on shell, we  have that the nucleon propagator
with momentum $p + q$ (this is part of the amputated amplitude
$t^{\mu\alpha}_A\Big|_{NP}$) can be placed on shell for a virtual
$W$. Thus, as discussed above in Eq.~(\ref{eq:avg}), we define the
tensor $B^{\nu\mu}$ as an angular average of the traces that appear
in the evaluation of the diagram. Namely,
\begin{equation}
B^{\nu\mu} =  \frac12 \int^{+1}_{-1} d\mu \frac1{2C_\rho} {\rm
Tr}\left\{\left(\slashchar{p}+M\right)\gamma^0 (t_A)^{\nu 
\dagger}_\alpha \gamma^0\left(\slashchar{p}+\slashchar{q}
-\slashchar{k}+M\right) (t_A)^{\mu\alpha} \right\}
\end{equation}
with $\mu = \vec{q}\cdot\vec{p}/|\vec{q}\,||\vec{p}\,|$. To simplify
the numerical integration, we have  given a small width
($\sim 10$ MeV) to the $p+q$ nucleon and have used an average for the
modulus of $\vec{p}$ and fixed the $(\vec{p}, \vec{q}\,)-$plane,
avoiding thus to perform two nested integrals. The total $W^+N \to N'
\rho$ amputated $t_A^{\mu\alpha}$ amplitude is obtained by summing
those corresponding to the $NP$, $CNP$, $\Delta P$ and $C\Delta P$
mechanisms. 

To deduce Eq.~(\ref{eq:W2p2h-t}), we have approximated the ph
$\rho-$selfenergy  (right-hand
part of the diagram depicted in Fig.~\ref{fig:rho-vt}) by
\begin{equation}
\Pi_{\alpha\beta} = \left(-g_{\alpha\beta} + k_\alpha k_\beta/k^2\right )
\hat\Pi^\lambda,\qquad  \hat\Pi^\lambda (k) = F_\rho^2(k) C_\rho
\frac{f^2_{\pi NN}}{m_\pi^2} \vec{k}_\pi^2   U_\lambda (k) \label{eq:rhpself}
\end{equation}
with $\lambda$ the charge of the $\rho-$meson. Eq.~(\ref{eq:rhpself})
is obtained by neglecting higher order terms, 
${\cal O} (\vec{l}^{\,2}/M^2)$, being
$\vec{l}= \vec{p}, \vec{q}$ or $\vec{k}$. This  is consistent with the
non-relativistic reduction that leads to the effective potentials in
Eqs.~(\ref{eq:st2x0}) and (\ref{eq:st2x3}).
\begin{figure}
\includegraphics[height=.56\textheight]{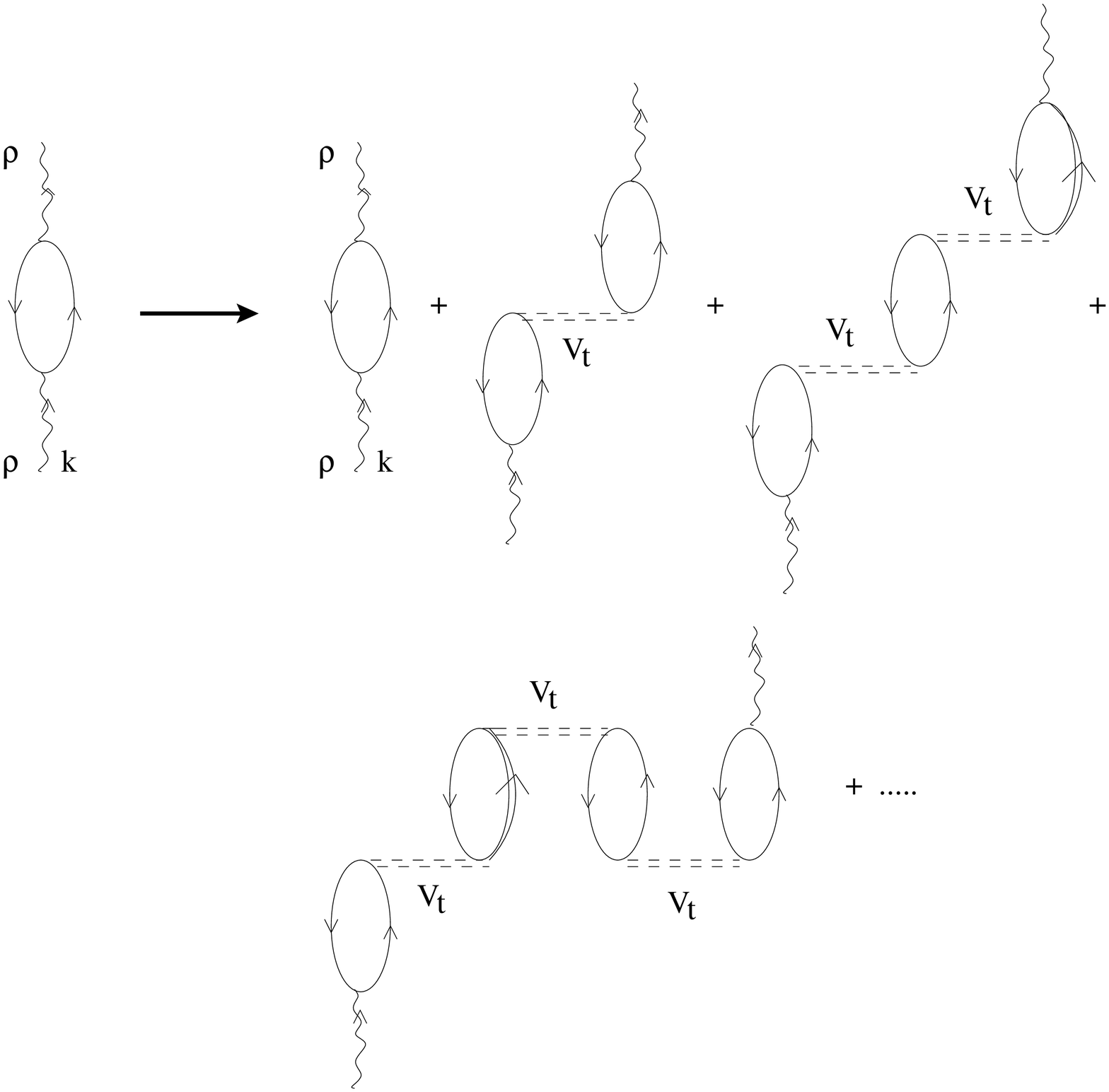}
 \caption {Diagrammatic representation of a series of RPA excitations
through ph and $\Delta$h excitations, driven by $V_t$.}
\label{fig:rpa-t}
\end{figure}

As previously done, we also
\begin{itemize}
\item  replace in Eq.~(\ref{eq:W2p2h-t}),
\begin{equation}
{\rm Im} U_\lambda (k) \Rightarrow \frac{{\rm Im} U_\lambda
  (k)}{|1-U(k) V_t|^2} \label{eq:rho-self}
\end{equation}
By including the non-relativistic Lindhard function for ${\rm
ph}\,+\,\Delta$h excitations in the denominator, we replace the ph
excitation of the right-hand in Fig.~\ref{fig:rho-vt} by a series of
RPA excitations through ph and $\Delta$h excitations, driven by $V_t$,
as depicted in Fig.~\ref{fig:rpa-t} (some more details will be given
below in the discussion of Eq.~(\ref{eq:ind})).

\item multiply $t_A^{\mu\alpha}$ by a factor 
\begin{equation}
t_A^{\mu\alpha} \Rightarrow t_A^{\mu\alpha} \times \left( 1+
\frac{g'_t}{F^2_\rho D_\rho C_\rho \vec{k}\,^2} \right)
\end{equation}
which allows us to replace the $\rho-$exchange interaction in
 Fig.~\ref{fig:rho-vt} by the  transverse part ($V_t$) of the
 effective ph--ph potential.
\end{itemize}
Finally, and to avoid double counting we must subtract the $NP$--$NP$
contribution from Eq.~(\ref{eq:W2p2h-t}), because this term was
already taken into account in the evaluation of the QE contribution,
through the inclusion of a realistic nucleon selfenergy, carried out
in Ref~\cite{Nieves:2004wx}.

\subsection{The $\Delta$ excitation term}
\label{sec:delta}

\begin{figure}
\makebox[0pt]{\includegraphics[height=.24\textheight]{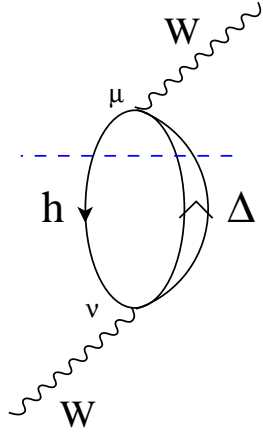}}
 \caption{Diagrammatic representation of the $\Delta$h
weak--nuclear excitation term.}
\label{fig:delta-term}
\end{figure}
Like in pion-nuclear, photo-nuclear and
electro-nuclear reactions at intermediate energies,  
the excitation of the $\Delta(1232)$ resonance by the weak probe is
expected to play a major role. This term is depicted in 
Fig.~\ref{fig:delta-term}, and its contribution to the
in medium $W^+-$selfenergy reads 
\begin{eqnarray}
-{\rm i} \Pi_{W;\Delta {\rm h}}^{\mu\nu}(q)&=& 
-2\cos^2\theta_C\left
(\frac{g}{2\sqrt 2} \right )^2\sum_N C_N^2 
 \int\frac{d^4p}{(2\pi)^4}  \frac{1}{p_\Delta^2-M^2_\Delta
+{\rm i} M_\Delta \Gamma_\Delta} {\rm
  i}\frac{\pi}{E(\vec{p}\,)}n_N(\vec{p}\,)\delta(p^0-E(\vec{p}\,))
 A^{\mu\nu}_\Delta(p,q)\,, \nonumber \\
A^{\mu\nu}_\Delta &=& \frac12 {\rm Tr}\left((\slashchar{p}+M)\gamma^0(\Gamma^{ \alpha \mu}
   )^\dagger\gamma^0P_{\alpha\beta} \Gamma^{\beta\nu}\right),
\label{eq:delta-contri}
\end{eqnarray}
with $p_\Delta = p+q$, $M_\Delta$ the resonance mass and
$\Gamma_\Delta$ its width, which can be found, e.g., in Eq.~(45) of
Ref.~\cite{Hernandez:2007qq}. The isospin factor $C_N$ takes the
values 1 and $\sqrt{3}$ for neutron and proton hole contributions,
respectively.  Finally, $P^{\mu\nu}(p_\Delta)$ is the spin 3/2
on-shell projection operator
\begin{equation}
P^{\mu\nu}(p_\Delta)= - (\slashchar{p}_\Delta + M_\Delta) \left [ g^{\mu\nu}-
  \frac13 \gamma^\mu\gamma^\nu-\frac23\frac{p_\Delta^\mu
  p_\Delta^\nu}{M_\Delta^2}+ \frac13\frac{p_\Delta^\mu
  \gamma^\nu-p_\Delta^\nu \gamma^\mu }{M_\Delta}\right]
\end{equation}
and
$\Gamma^{\alpha\nu}(p,q)$ is the weak $N\Delta$ transition vertex, that can be found in
Eq. (40) of Ref.~\cite{Hernandez:2007qq}.  The 
contribution to the hadron tensor from the selfenergy of
Eq.~(\ref{eq:delta-contri}) is\footnote{Note that the tensor
$A^{\mu\nu}_\Delta$ can be split as in Eq.~(\ref{eq:sym}).}
\begin{equation}
 W_{\Delta {\rm h}}^{\mu\nu}(q) = -2 \cos^2\theta_C \Theta(q^0) \sum_N C_N^2  
\int \frac{d^3 r}{2\pi}\frac{d^3p}{(2\pi)^3}
\frac{n_N(\vec{p}\,)}{E(\vec{p}\,)}  {\rm Im}\left( \frac{1}{p_\Delta^2-M^2_\Delta
+{\rm i} M_\Delta \Gamma_\Delta}\right) A^{\nu\mu}_\Delta
\Big|_{p^0=E(\vec{p}\,)} \label{eq:Wdelta-contri}
\end{equation}
\begin{figure}
\includegraphics[height=.36\textheight]{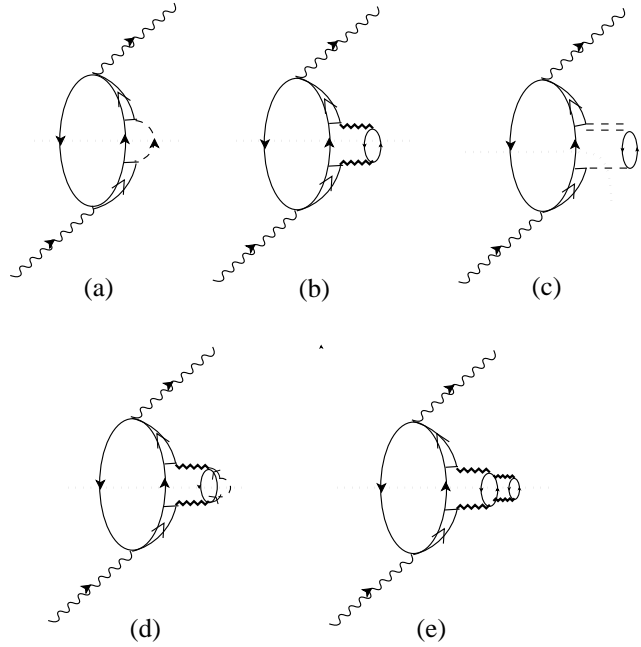}
 \caption {Diagrammatic representation of the different contributions
   of the $\Delta $h weak--nuclear excitation term.}
\label{fig:delta}
\end{figure}
Some remarks are here in order. For instance, we must be careful to
avoid double counting. Indeed, the contribution to the hadron tensor
of the $\Delta $h excitation term arises from the imaginary part of
the $\Delta$ propagator, and in particular from the
$\Delta-$width. One of the terms implicit in Eq.~(\ref{eq:W1pi}), the
one where one picks up the $\Delta$ excitation term both in $j^\nu_A$
and in $j^{\dagger \mu}_A$ (depicted diagrammatically in
Fig.~\ref{fig:delta}(a)), gives precisely the same contribution plus
some medium corrections that take into account the Pauli blocking
effects. Thus, if we would naively add it to the hadronic tensor, the
contribution of Eq.~(\ref{eq:Wdelta-contri}) would be counted twice.
Indeed, the term of Fig.~\ref{fig:delta}(a) can be cast in the form of
the diagram of Fig.~\ref{fig:delta-term}, but with a
$\Delta-$selfenergy insertion constructed from a pion loop.  When the
pion is put on the mass shell to build the hadron tensor, we obtain
the $\Delta-$width and thus qualitatively the equivalence is
shown\footnote{More quantitatively, to evaluate this term we can start
from Eq.~(\ref{eq:W1pibis}) and perform the $d^4 k_\pi$ integration in
order to show explicitly the $1\pi$ $\Delta-$selfenergy
$\Sigma^{\alpha\beta}_\pi$.  We get
\begin{eqnarray}
-{\rm i} \Pi_{W;1{\rm ph}1\pi}^{\mu\nu}\Big|_{\Delta}(q)&=& 
-2 {\rm i} \cos^2\theta_C\left
(\frac{g}{2\sqrt 2} \right )^2\sum_N C_N^2 
 \int\frac{d^4p}{(2\pi)^4} 
 \frac{G(p;\rho_N)}{|(p+q)^2-M^2_\Delta+{\rm i} M_\Delta
   \Gamma_\Delta|^2} \times \nonumber \\
& \times & \frac12 {\rm Tr}\left((\slashchar{p}+M)\gamma^0(\Gamma^{ \alpha \mu}
   )^\dagger\gamma^0P_{\alpha\beta} ({\rm -i}\Sigma^{\beta
\delta}_{\Delta;\pi} )
 P_{\delta\epsilon}\Gamma^{\epsilon\nu} \right)  \label{eq:delta1} \\
{\rm -i}\Sigma^{\beta \delta}_{\Delta;\pi} (p_\Delta) &=&
-\left(\frac{f^*_{\pi N\Delta}}{m_\pi}\right)^2
\int \frac{d^4k_\pi}{(2\pi)^4} D_\pi(k_\pi) k^\beta_\pi k^\delta_\pi
G(p+q-k_\pi; \rho)
\left (\slashchar{p}+ \slashchar{q}-\slashchar{k_\pi}+ M\right)\,
\end{eqnarray}
where $f^*_{\pi N\Delta}=2.14$ is the $\pi N \Delta$ coupling, and for
simplicity we have evaluated the $\Delta-$selfenergy for a symmetric
Fermi sea. In principle, the $\Delta-$selfenergy $\Sigma^{\beta
\delta}_{\Delta;\pi}$, which is a matrix in the Dirac space and a
Lorentz tensor can be expressed in terms of a linear combination of
the five orthogonal spin projection operators introduced in Eq.~(10)
of Ref.~\cite{Benmerrouche:1989uc}. The coefficients, $A_i$ of such
linear combination will be matrices in the Dirac space and Lorentz
scalars. We will enormously simplify the discussion here neglecting
$\Delta-$offshell effects. Within this approximation, the spin 3/2
projector of Ref.~\cite{Benmerrouche:1989uc} reduces to that used here
($-P^{\mu\nu}/2M_\Delta$) to construct the $\Delta-$propagator which
satisfies $P^{\mu\nu}P_{\nu\delta}=-2M_\Delta P^\mu_\delta $ and
$\slashchar{p}_\Delta P^{\mu\nu}= P^{\mu\nu} \slashchar{p}_\Delta =
M_\Delta P^{\mu\nu }$.  Since in Eq.~(\ref{eq:delta1}), 
$\Sigma_{\Delta;\pi}^{\beta \delta}$ always appears contracted with
two projector operators, one realizes that using the orthogonality
properties, only the scalar quantity $\Sigma_{\Delta;\pi}$, defined as
\begin{equation}
\Sigma_{\Delta;\pi}^{\beta \delta} = - \Sigma_{\Delta;\pi} P^{\beta \delta} +
\underbrace{\cdots}_{\perp P^{\beta \delta}  }
\end{equation}
will contribute in  Eq.~(\ref{eq:delta1}), which now gets simplified to
\begin{eqnarray}
-{\rm i} \Pi_{W;1{\rm ph}1\pi}^{\mu\nu}\Big|_{\Delta}(q)&=& 
2\cos^2\theta_C\left
(\frac{g}{2\sqrt 2} \right )^2\sum_N C_N^2 
 \int\frac{d^4p}{(2\pi)^4} G(p;\rho_N)
 \frac{4M^2_\Delta \Sigma_{\Delta;\pi}}{|(p+q)^2-M^2_\Delta+{\rm i} M_\Delta
   \Gamma_\Delta|^2}  A^{\mu\nu}_\Delta(p,q) \label{eq:delta-aux}
\end{eqnarray}
Thus, we recover Eq.~(\ref{eq:Wdelta-contri}) from
Eq.~(\ref{eq:delta-aux}) when we replace $G(p; \rho_N)$ by ${\rm
i}\frac{\pi}{E(\vec{p}\,)}n_N(\vec{p}\,)\delta(p^0-E(\vec{p}\,))$ to
get rid of the vacuum contribution, and consider the imaginary part of
the $\Delta-$ selfenergy in Eq.~(\ref{eq:delta-aux}). This is because
to obtain the hadron tensor it appears always ${\rm Im}
\Sigma_{\Delta;\pi}$, thanks to the symmetry properties of the tensor
$ A^{\mu\nu}_\Delta$, and by noting that
\begin{equation}
{\rm Im} \Sigma_{\Delta;\pi} = \frac{\Gamma_\Delta}{4 M_\Delta},
\end{equation}
up to density corrections which will account for Pauli
blocking. This 
latter relation follows from the Dyson equation for the $\Delta$
propagator, within the on-shell approximation we are using,
\begin{equation}
 \frac{ {\rm i} P^{\mu\nu}}{p^2_\Delta-M^2_\Delta} + 
\frac{{\rm i}P_{\mu\beta}}{p^2_\Delta-M^2_\Delta} \left(-{\rm
  i}\Sigma_{\Delta;\pi}^{\beta\delta}  \right)   
\frac{{\rm i} P_{\delta\nu}}{p^2_\Delta-M^2_\Delta} +\cdots = 
\frac{{\rm i}P^{\mu\nu}}{p^2_\Delta-M^2_\Delta+4M^2_\Delta \Sigma_{\Delta;\pi}}  
\end{equation}
}.
In a similar way, the diagram of Fig.~\ref{fig:delta}(b) is one of the
terms implicit in the diagram of Fig.~\ref{fig:2ph} that produces a
2p2h excitation.

However, given the importance of the $\Delta-$pole contribution and
since the $\Delta$ properties are strongly modified inside the nuclear
medium~\cite{Gil:1997bm,Hirata:1978wp,Oset:1981ih,Oset:1987re,Nieves:1993ev,AlvarezRuso:2007tt,Amaro:2008hd},
a more careful treatment of the $\Delta$ mechanisms is advisable. This
implies some additional nuclear corrections to
Eq.~(\ref{eq:Wdelta-contri}) to include the full effect of the
self-energy of the $\Delta$ in the medium
$\Sigma_\Delta(\rho(\vec{r}\,))$ in a systematic manner.  In addition,
these corrections provide genuinely new contributions to the hadronic
tensor (e.g. 3p3h mechanisms).  Here, we follow the same approach as
in Ref.~\cite{Gil:1997bm}, which is based on
Refs.~\cite{Oset:1987re,Nieves:1993ev,Nieves:1991ye}.  In the nuclear
medium the resonance self-energy is modified because of several
effects such as Pauli blocking of the final nucleon and absorption
processes: $\Delta N \to NN$, $\Delta N \to NN\pi$, or $\Delta NN \to
NNN$.  This is done using a ph--ph interaction that includes, besides
pion exchange, SRC, a transverse channel driven by $\rho-$exchange
(see Eq.~(\ref{eq:veff})) and a RPA-re-summation.

\begin{figure}
\includegraphics[height=.28\textheight]{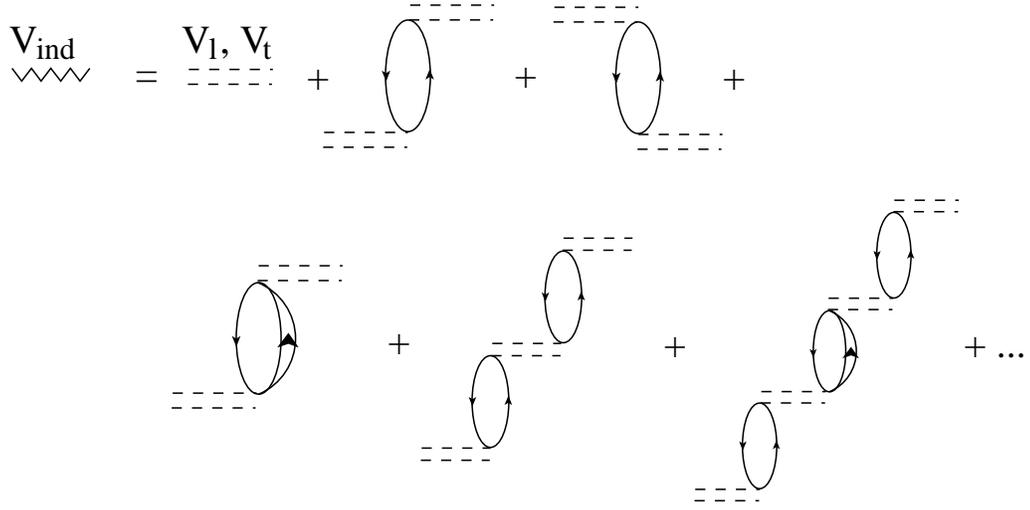}
 \caption  {Diagrammatic representation of the induced interaction.}
\label{fig:induced}
\end{figure}

Following this approach,
in the $\Delta-$propagator, we  approximate
\begin{equation}
\frac{1}{p_\Delta^2-M^2_\Delta
+{\rm i} M_\Delta \Gamma_\Delta} \sim
\frac{1}{\sqrt{s}+M_\Delta}\frac{1}{\sqrt{s}-M_\Delta+ {\rm i} 
 \Gamma_\Delta/2} \label{eq:replacements}
\end{equation}
with $s=p_\Delta^2$. In the particle propagator of the right hand side
of the above equation, we make the substitution: $\Gamma_\Delta/2 \to
\Gamma_\Delta^{\rm Pauli}/2- {\rm Im} \Sigma_\Delta$ and take ${\rm
Im} \Sigma_\Delta(\rho(\vec{r}\,))$ and $\Gamma_\Delta^{\rm Pauli}/2$
as follows.  First, the Pauli blocking\footnote{In the diagram
(a) of Fig.~\ref{fig:delta} appears the factor
$n_N(\vec{p}\,)(1-n_{N'}(\vec{p}+\vec{q}-\vec{k}_\pi))$, see
Eq.~(\ref{eq:W1pi}).} of the $\pi N$ decay 
 reduces the $\Gamma_\Delta$ free width to $\Gamma_\Delta^{\rm
   Pauli}$, which can be found in Eq.~(15) of
Ref.~\cite{Nieves:1991ye}. Next, the imaginary part of the selfenergy
in Eq.~(\ref{eq:replacements}) accounts for the diagrams depicted in
Fig.~\ref{fig:delta}, where the double dashed line stands for the
effective spin--isospin interaction, while the wavy line accounts
for the induced interaction.  The effective spin-isospin interaction
is originated by  $\pi$ and $\rho$ exchange in the presence of short range
correlations. It is obtained by substituting~\cite{Oset:1987re,Nieves:1993ev,Nieves:1991ye}
\begin{equation}
\left(\frac{f_{\pi NN}}{m_{\pi}}\right)^2\hat{k}^\pi_i\hat{k}^\pi_
 j \vec{k}_\pi^{\,2} D_{\pi}(k_\pi)
 \rightarrow \hat{k}^\pi_i\hat{k}^\pi_j V_{l}(k_\pi) +(\delta_{ij}-
 \hat{k}^\pi_i\hat{k}^\pi_j)V_{t}(k_\pi) \label{eq:veff}
\end{equation}
with $\hat{k}^\pi_i= k^\pi_i/|\vec{k}_\pi|$. The induced interaction
accounts for the series of diagrams depicted in
Fig.~\ref{fig:induced}. There is an RPA sum through particle--hole and
$\Delta$h excitations and it is readily obtained as
\begin{eqnarray}
V_{\rm ind}&=&
\hat{k}^\pi_i\hat{k}^\pi_j\,
\frac{V_l(k_\pi)}{1-U(k_\pi)V_l(k_\pi)}
+
(\delta_{ij}-\hat{k}^\pi_i\hat{k}^\pi_j)
\frac{V_t(k_\pi)}{1-U(k_\pi)V_t(k_\pi)} \label{eq:ind}
\end{eqnarray}
where $U (k_\pi)$ is the non-relativistic Lindhard function for
${\rm ph}\,+\,\Delta $h excitations\footnote{The different couplings
for $N$ and $\Delta$ are incorporated in $U_N$ and $U_\Delta$ and then
the same interaction strengths $V_l$ and $V_t$ are used for ph and
$\Delta $h excitations~\cite{Oset:1981ih}. } (see
Eq.~(\ref{eq:def_lin})).  The evaluation of $\Sigma_\Delta$ is done in
Ref.~\cite{Oset:1987re}. The imaginary part of $\Sigma_\Delta$ can be
parametrized as
\begin{equation}
- {\rm Im} \Sigma_\Delta(\rho(\vec{r}\,)) = C_Q\left ( 
\frac{\rho}{\rho_0}  \right)^\alpha + C_{A_2} \left ( 
\frac{\rho}{\rho_0}  \right)^\beta +  C_{A_3} \left ( 
\frac{\rho}{\rho_0}  \right)^\gamma \label{eq:imsigma}
\end{equation}
where the different coefficients are given\footnote{The
  parameterizations are given as a function
of the kinetic energy in the laboratory system of a pion that would
excite a $\Delta$ with the corresponding invariant mass and are valid
in the range 85 MeV $ < T_\pi < 315$ MeV. Below 85 MeV the
contributions from $C_Q$ and $C_{A_3}$ are rather small and are taken
from Ref.~\cite{Nieves:1991ye}, where the model was extended  to low
energies. The term with $C_{A_2}$ shows a very mild energy dependence
and we still use the parametrization from Ref.~\cite{Oset:1987re} even
at low energies. For $T_\pi$ above 315 MeV we have kept these
self-energy terms constant and equal to their values at the bound. The
uncertainties in these pieces are not very relevant there because the
$\Delta \to N\pi$ decay becomes very large and absolutely dominant.} 
in Eq.(4.5) and Table 2 of 
Ref.~\cite{Oset:1987re}.  The separation of terms in
Eq.~(\ref{eq:imsigma}) is useful because the term $C_Q$ comes from the
$\Delta N \to NN\pi$ process (diagrams (c) and (d) of
Fig.~\ref{fig:delta} when the lines cut by the dotted line are placed
on shell, and hence the term is related to the $( W^*, \pi)$
channel), while $C_{A2}, C_{A3}$ come from diagrams (b) and (e) and
are related to two ($W^* N N \to NN$) and three ($W^*NNN \to NNN$)
body absorption. Hence, the separation in this formula allows us to
separate the final cross section into different channels.

To avoid double counting, we must subtract the contribution of the
 $\Delta P$--$\Delta P$ diagram of Fig.~\ref{fig:delta}(a) from
 Eq.~(\ref{eq:1ph1pi}), already taken into account through the
 $\Gamma_\Delta^{\rm Pauli}$ piece of the $\Delta$ selfenergy. We must
 also subtract the contribution of the $\Delta P$--$\Delta P$ diagram
 of Fig.~\ref{fig:delta}(b) from Eqs.~(\ref{eq:W2p2h}) and
 (\ref{eq:W2p2h-t}), because these terms were already taken into
 account in the evaluation of the $C_{A_2}$ contribution to the
 $\Delta-$selfenergy~\cite{Oset:1987re}.

\begin{figure}
\includegraphics[height=.2\textheight]{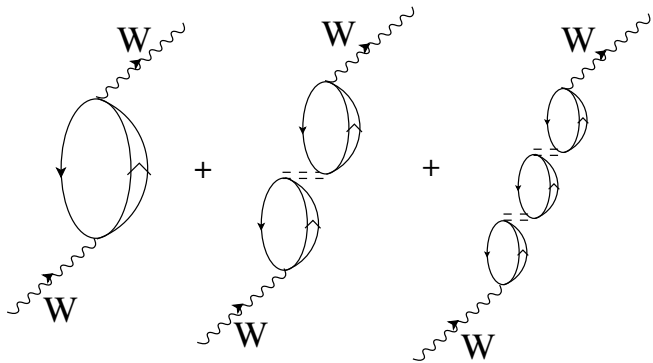}
 \caption  {Irreducible pieces in the $\Delta$h channel from the
 $\Delta$h interaction.}
\label{fig:rpa}
\end{figure}

To end this subsection, we would like to devote a few words to the
real part of the $\Delta-$selfenergy and the RPA sum of $\Delta$h
excitations shown in Fig.~\ref{fig:rpa}. Both of them produce effects
on the nuclear response to the weak probe that partially cancel
out. In Ref.~\cite{Oset:1987re}, the dispersive contributions to ${\rm
Re}\Sigma_\Delta$ associated to the diagrams that gave rise to ${\rm
Im}\Sigma_\Delta$ were also computed. There, it was found ${\rm
Re}\Sigma^{(0)}_\Delta \sim -50 \rho/\rho_0$ [MeV] at $T_\pi = 50$ MeV
and a smooth dependence on the pion energy. In principle, ${\rm
Re}\Sigma^{(0)}_\Delta$ could be taken into account by substituting
$M_\Delta \to M_\Delta + {\rm Re}\Sigma^{(0)}_\Delta$ in the particle
propagator of the right hand side of Eq.~(\ref{eq:replacements}). On
the other hand, it is  easy to realize that the RPA sum of $\Delta
$h excitations, shown in Fig.~\ref{fig:rpa} can be  cast as a
contribution to the real part of the
$\Delta-$selfenergy~\cite{Gil:1997bm}. Actually, the
latter depends on the particular component of the hadron tensor
$W^{\mu\nu}_{\Delta {\rm h}}$ which is being evaluated . Thus, for
instance, the RPA series depicted in Fig.~\ref{fig:rpa} can be taken
into account, when computing $W^{xx}_{\Delta {\rm h}}$ or
$W^{yy}_{\Delta {\rm h}}$ (transverse components to the direction of
the $W-$boson) by replacing ${\rm Re}\Sigma^{(0)}_\Delta$ by ${\rm
Re}\Sigma^{(0)}_\Delta +4 \rho V_t /9$. This latter sum, in good
approximation, is positive for the whole range of energies studied
here. This was the situation for the inclusive $(e,e')$ nuclear
reaction studied in Ref.~\cite{Gil:1997bm}, since there, the
excitation of the $\Delta$ resonance by the virtual photon selected
the transverse mode of the RPA series (see discussion of Eq.~(44) in
Ref.~\cite{Gil:1997bm}).  However, when the longitudinal component
$W^{zz}_{\Delta {\rm h}}$ is evaluated, the longitudinal part, $V_l$,
of the effective spin--isospin interaction is selected and now, this
RPA sum is taken into account by substituting\footnote{Note that, in
the studies of neutrino induced pion coherent production in nuclei
carried out in Refs.~\cite{AlvarezRuso:2007tt,Amaro:2008hd}, the
replacement ${\rm Re}\Sigma^{(0)}_\Delta$ by ${\rm
Re}\Sigma^{(0)}_\Delta +4 \rho g' /9$ is employed to account for the
corresponding RPA re-summations.  The Landau Migdal parameter $g'$
used there is part of $V_l$, which in addition also includes explicit
pion-exchange (see
Eq.~(\ref{eq:st2x0}))~\cite{Oset:1987re,Nieves:1993ev,Nieves:1991ye}. This
latter contribution was not considered in the works of
Refs.~\cite{AlvarezRuso:2007tt,Amaro:2008hd, AlvarezRuso:2007it},
because there the distortion of the pion, by using an outgoing
solution of the Klein-Gordon equation with the optical pion--nucleus
potential derived in Ref.~\cite{Nieves:1991ye}, was implemented, and
it accounts for the RPA-renormalization induced by the
$\Delta$h--$\Delta$h pion-exchange interaction.  } ${\rm
Re}\Sigma^{(0)}_\Delta$ by ${\rm Re}\Sigma^{(0)}_\Delta +4 \rho V_l
/9$, which shows a more pronounced $q^2$ dependence than the
combination that appeared in the RPA renormalization of transverse
components of the hadronic tensor. Indeed, it turns out that the ${\rm
Re}\Sigma^{(0)}_\Delta +4 \rho V_l /9$ combination does not have a
well defined sign for the whole kinematical range of energies studied
in this work. Setting to $M_\Delta$ the position of the pole of the
$\Delta$ propagator, or changing it by adding or subtracting to
$M_\Delta$ about 30 MeV, as it could be inferred from the typical
values that ${\rm Re}\Sigma^{(0)}_\Delta +4 \rho V_{l(t)} /9$ takes
for the relevant kinematics to this work, leads to trivial shifts in
the position of the $\Delta-$peak, moderately changes of the strength
(around 20 \%) at the maximum and very tiny changes of the
$q^0-$differential shape. Of course, all these effects induced by the
RPA--re-summation might be properly taken into account, as it was done
for the case of the QE-region\footnote{For QE kinematics, taking into
account properly the RPA effects is much more
important~\cite{Nieves:2004wx} than in the $\Delta-$region, since the
cancellation of their effects with the difference between particle and
hole selfenergies is much less effective.} in
Ref.~\cite{Nieves:2004wx}, but they, in conjunction with ${\rm
Re}\Sigma^{(0)}_\Delta$, would induce changes smaller than both, the
precision in the current experimental determination of cross sections,
and the uncertainties due to our lack of a precise knowledge of the
axial nucleon-to-$\Delta$ transition form factor
$C_5^A$~\cite{Hernandez:2010bx}. For simplicity, in this work we will
not renormalize the real part of the position of the $\Delta-$peak,
which eventually could be studied in the future when more accurate
measurements become available.

\subsection {CC  antineutrino induced reactions}
\label{sec:anti}
The cross section for  the antineutrino induced nuclear reaction
\begin{equation}
{\bar \nu}_l (k) +\, A_Z \to l^+ (k^\prime) + X  \label{eq:anti}
\end{equation}
is easily obtained from the expressions given in the previous
subsections, by changing the sign of the antisymmetric part of the
lepton tensor ($L^a$) and using the $W^- N \to N^\prime \pi^\lambda$
amplitudes of Ref.~\cite{Hernandez:2007qq}, instead of those involving
the $W^+$ gauge boson. Note that the pion production off the nucleon
amplitudes give rise, directly on indirectly, to all contributions
considered here, except the QE ones. We take the 
$\bar \nu-$QE cross sections from  Ref.~\cite{Nieves:2004wx}.

\section{Results}
\label{sec:res}
\begin{figure}
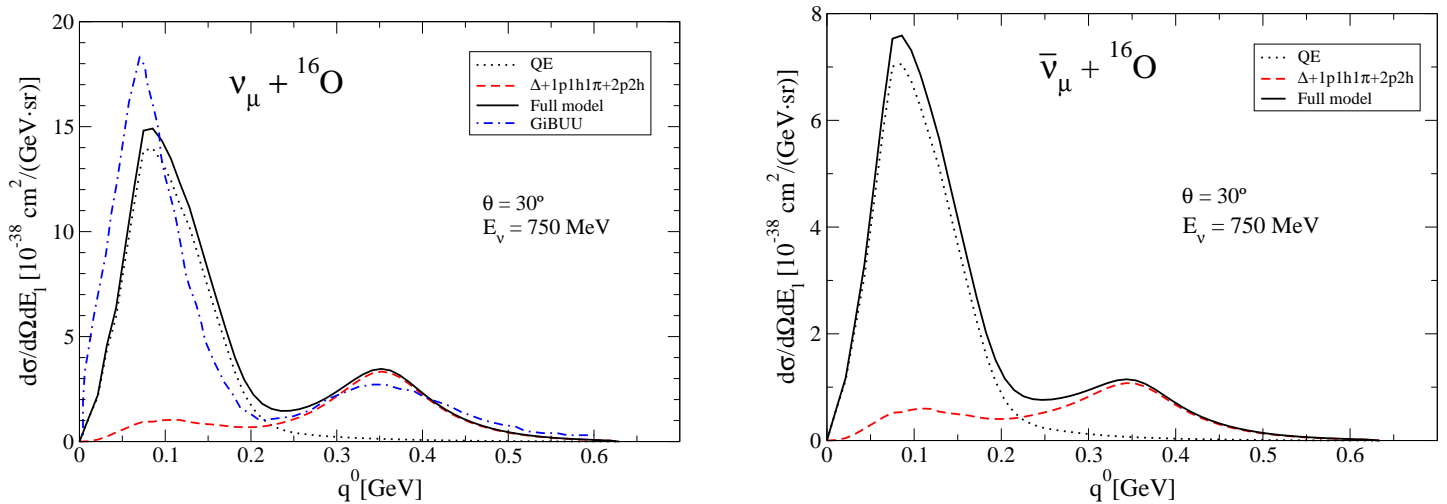

\makebox[0pt]{\includegraphics[width=0.5\textwidth]{30deg750MeV.eps}
\hspace{1cm}\includegraphics[width=0.5\textwidth]{anti-30deg750MeV.eps}}
\caption{Muon neutrino (left) and antineutrino (right) CC differential
cross section $\frac{d\sigma}{d\Omega dE_\mu}$ in oxygen, at 30 degrees of
scattering angle  and with an incident neutrino energy of
750 MeV, plotted against the transferred energy to the
nucleus. Different contributions are displayed, standing the solid
lines for our full model results. Besides in the left panel, we also
show results (blue dash-dotted line) from
Ref.~\cite{Leitner:2008ue} and obtained within the GiBUU framework.}
\label{fig:30deg750MeV}
\end{figure}

\begin{figure}
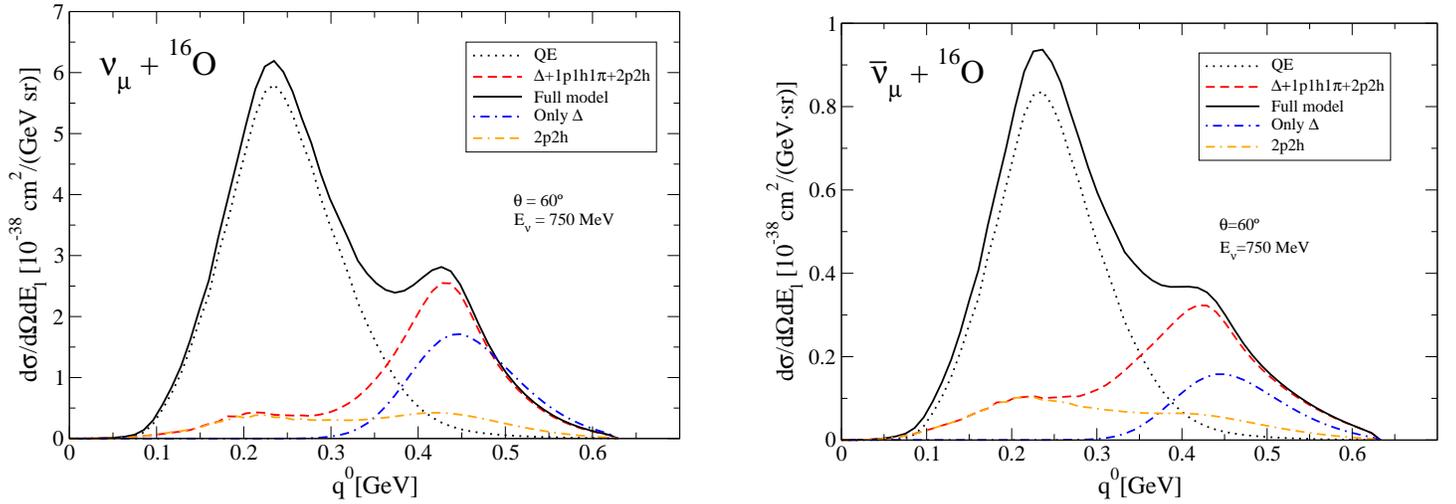

\makebox[0pt]{\includegraphics[width=0.5\textwidth]{60deg750MeV2p2h.eps}
\hspace{1cm}\includegraphics[width=0.5\textwidth]{anti-60deg750MeV.eps}}
\caption{Muon neutrino (left) and antineutrino (right) CC differential
cross section $\frac{d\sigma}{d\Omega dE_\mu}$ in oxygen, at 60 degrees of
scattering angle  and with an incident neutrino energy of
750 MeV, plotted against the transferred energy to the
nucleus. The solid
lines stand for our full model results.}
\label{fig:60deg750MeV}
\end{figure}

We will mainly focus here in the dip and $\Delta-$peak regions, since
the QE contribution was discussed at length in
Ref.~\cite{Nieves:2004wx}. In Fig.~\ref{fig:30deg750MeV}, we show
results for both muon neutrino (left) and antineutrino (right) induced
CC differential cross sections at 30 degrees as a function of the
energy transferred to the nucleus ($^{16}$O). The incoming neutrino
(anti neutrino) energy is 750 MeV. We clearly observe both the
$\Delta(1232)$ and the QE peaks; for this scattering angle, the QE
contribution turns out to be significantly larger than that of the
$\Delta$ resonance. We split the full contribution into the QE and non
QE ($\Delta+$1p1h1$\pi$+ 2p2h) parts. General features are the same
for both neutrino and antineutrino induced cross sections, and the
main difference is an homogeneous reduction in the size of the
differential cross section.  For comparison, in the left panel (blue
dashed-dotted line) we also display some results from
Ref.~\cite{Leitner:2008ue}, obtained within the Giessen
Boltzmann-Uehling-Uhlenbeck (GiBUU) framework, which takes into
account various nuclear effects: the local density approximation for
the nuclear ground state, mean-field potentials, and in-medium
spectral functions, but does not include those due to RPA
correlations. We note first, some discrepancies between these results
and ours in the QE region, which origin can be traced back to the
implementation of RPA corrections in our
scheme~\cite{Nieves:2004wx}. Indeed, the found differences (small
shift in the position and reduction in size, about 25\%, of the QE
peak) are qualitatively identical to those existing between data and
GiBUU predictions for the case of inclusive electron cross section for
a similar kinematics (incident electron energy of 700 MeV and
scattering angle of 32 degrees) showed in the upper panel of
Fig.~9 of Ref.~\cite{Leitner:2008ue}. On the other hand, in this
latter figure can be also appreciated the differences with the GiBUU
model in the description of the dip region. Indeed, we see in
Fig.~\ref{fig:30deg750MeV} that in the dip region, our model predicts
larger cross sections than those obtained within the GiBUU
scheme. This is due to the 2p2h mechanisms of Figs.~\ref{fig:2ph} and
\ref{fig:rho-vt} included in our model. Actually, these contributions
make also our cross section at the $\Delta-$peak larger than the one
predicted in Ref.~\cite{Leitner:2008ue}, even though we use a value of
$C_5^A(0)$ smaller than that used in Ref.~\cite{Leitner:2008ue} (1 vs
1.2). For larger scattering angles, the dip-region cross section
becomes relatively much more important, and thus the inclusion of the
2p2h contributions turns out to be of larger relevance. This is
clearly appreciated in Fig.~\ref{fig:60deg750MeV}, where we show
results at 60 degrees. In this figure, besides the separation between
QE and non QE contributions to the differential cross section, the
2p2h part\footnote{A small three body absorption (3p3h) contribution,
induced by $C_{A_3}$ in the $\Delta-$selfenergy of
Eq.~(\ref{eq:imsigma}) is also included under the label 2p2h in
Fig.~\ref{fig:60deg750MeV}, and in what follows. } of this latter
contribution is shown (orange double dash-dotted curves). The blue
dash-dotted lines stand in this figure for the results obtained from
only the $\Delta$h weak--nuclear excitation term of
Fig.~\ref{fig:delta}(a), neglecting Pauli blocking effects affecting
the in medium resonance width. We see how the systematic many body $W-$absorption
modes and the in medium effects considered here change drastically the
nuclear response function in the $\Delta-$peak, as happened in the QE
region as well~\cite{Nieves:2004wx}.

The 2p2h cross section accounts for events were the gauge boson is
absorbed by a pair of nucleons, in contrast to QE events for which it
is absorbed by one nucleon, and furthermore no pions are being
produced in this first step. Up to re-scattering processes which could
eventually produce secondary pions, 2p2h events will give rise to
only one muon to be detected. Thus, they could be experimentally
misidentified as QE events. Yet, 1p1h1$\pi$ events, in which the
resulting pion from the $W$ absorption is subsequently absorbed and
does not come off the nucleus, could be also misinterpreted as QE
events, if only leptons are being detected. A correct identification
of CCQE events, which is the signal channel in oscillation
experiments, is relevant for neutrino energy reconstruction and thus
for the oscillation result.  By looking at the 2p2h contribution in
Fig.~\ref{fig:60deg750MeV}, we see that at least about 15\% of the
quasielastic cross section might be misidentified in present-day
experiments and need to be corrected for by means of event
generators. As mentioned above, 1p1h1$\pi$ mechanisms followed by the
absorption of the resulting pion, will even make worse the
situation~\cite{Leitner:2010kp}.

\begin{figure}[t]
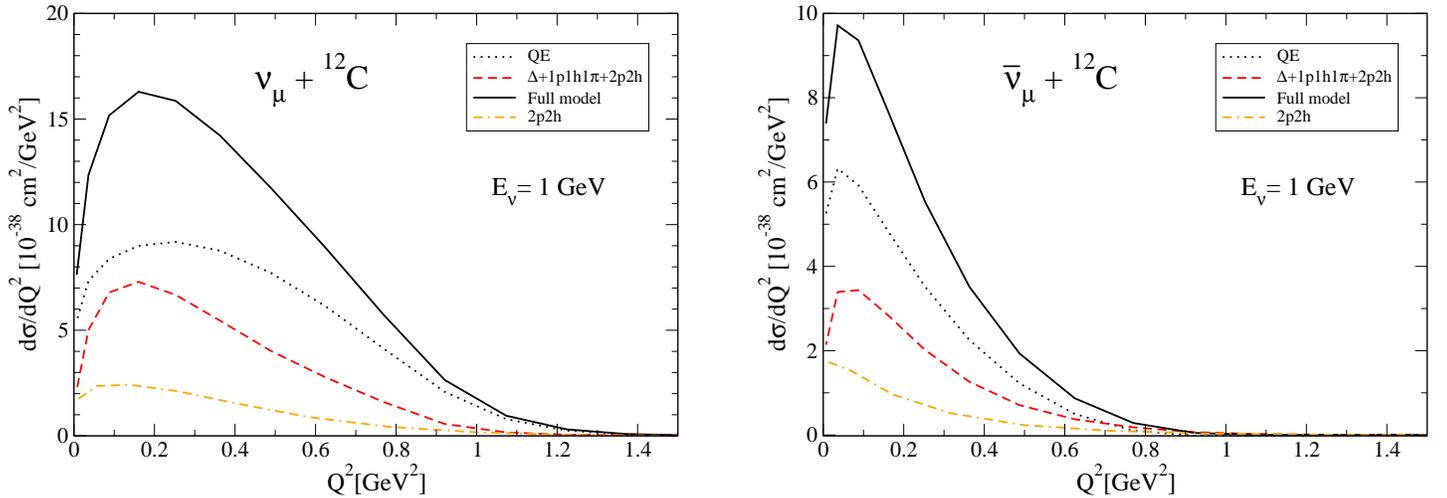

\makebox[0pt]{\includegraphics[width=0.5\textwidth]{Q2distributions.eps}\hspace{1cm}\includegraphics[width=0.5\textwidth]{Q2distributions_antineutrino.eps}} 
\caption{Muon neutrino (left) and antineutrino (right) CC differential
cross section $\frac{d\sigma}{dQ^2}$ in carbon for 
an incident neutrino energy of
1 GeV ($Q^2=-q^2)$. Different contributions are displayed, standing the solid
lines for our full model results.}
\label{fig:q21000MeV}
\end{figure}

In Fig.~\ref{fig:q21000MeV}, we show CC $q^2$ differential cross
sections in carbon for an incident energy of 1 GeV. We observe that
the 2p2h contribution is sizeable for both, neutrino and
antineutrino induced reactions, and that it shows a less pronounced $q^2$
dependence than the QE or the $\Delta+1p1h1\pi$ components of the total
result. On the other hand, the antineutrino distribution is much 
narrower than the neutrino one. Neglecting lepton mass effects, 
both distributions should be equal at $q^2=0$, and since the antineutrino
cross sections are smaller than the neutrino ones, is reasonable to expect
 the  $\nu$ distributions to be wider than  the $\bar\nu$ ones.

\begin{figure}
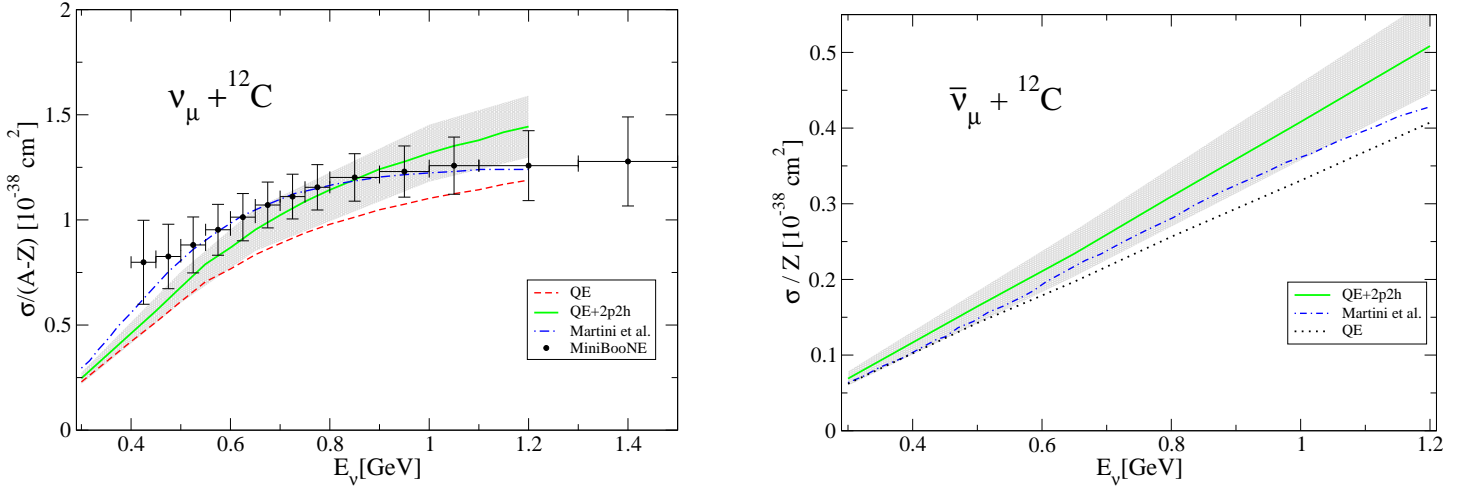

\makebox[0pt]{\includegraphics[width=0.5\textwidth]{MiniBooNEQE.eps}
\hspace{1cm}
\includegraphics[width=0.5\textwidth]{MiniBooNEQE_antineutrino.eps}}
\caption{Left: Flux-unfolded MiniBooNE $\nu_\mu$ CCQE cross section
  per neutron as a function of neutrino energy (data-points) from
  Ref.~\cite{AguilarArevalo:2010zc}, together with different
  theoretical predictions from this work. Right: Different theoretical
  calculations for antineutrino cross sections per proton off $^{12}$C
  as a function of the antineutrino energy. For comparison, in both
  plots, we also show the results (blue dash-dotted line) of Martini and
  collaborators taken from  Ref.~\cite{Martini:2010ex}. Bands accounting
  for the theoretical errors affecting our results are displayed in
  both panels.}\label{fig:sQE}
\end{figure}

The MiniBooNE collaboration has measured~\cite{AguilarArevalo:2010zc}
the muon neutrino's CCQE cross section on $^{12}$C. The flux-unfolded
results as a function of the neutrino energy are depicted in the left
panel of Fig.~\ref{fig:sQE}, together with different predictions from
the scheme presented here.  The first observation is that our QE curve
misses the data-points, being our predicted QE cross section
significantly smaller than those reported by the MiniBooNE
collaboration. Actually in ~\cite{AguilarArevalo:2010zc}, and to
achieve a reasonable description of the data, an unexpectedly high
effective axial mass $M_A^{\rm eff}$ (entering in the axial-vector
$WNN$ form-factor) of 1.35 GeV had to be used in the relativistic FG
model implemented in the NUANCE event generator employed by the
MiniBooNE collaboration. This value of $M_A$ is significantly larger
than the world average value $M_A=1.03$ GeV. It is interesting to
note, however, that in Ref.~\cite{AguilarArevalo:2010zc} is also
pointed out the NOMAD~\cite{Lyubushkin:2008pe} and
LSND~\cite{Auerbach:2002iy} high energy ($E_\nu > 4$ GeV) CCQE cross
sections are better described with the world average value for
$M_A$. The situation become even more worrying, after the work of
Ref.~\cite{Benhar:2010nx}.  That work finds that a 
theoretical approach based on the impulse approximation and realistic
spectral functions, successfully applied to QE electron scattering,
fails  to reproduce the CCQE neutrino-nucleus cross section,
unless the value of the nucleon axial mass resulting from deuteron
measurements is significantly increased. In addition, they also rule
out the possibility, advocated in Ref.~\cite{AguilarArevalo:2010zc},
of interpreting the large $M_A$ resulting from the MiniBooNE analysis
as an effective axial mass, modified by nuclear effects beyond the FG
model~\cite{Benhar:2009wi}. Actually, in \cite{Benhar:2010nx}, it is
 suggested that the  many body techniques successfully applied
in QE electron-nucleus scattering are not able to explain neutrino
induced cross sections and it is  argued that the development of a
new {\it paradigm}, suitable for application to processes in which the
lepton kinematics is not fully determined, will be required.

Our results do not support this last statement/interpretation, and we
rather agree with the picture that emerges from the works of
M. Martini et al.~\cite{Martini:2009uj,Martini:2010ex}. These latter
works, in our opinion, constituted a significant step forward to
clarify the situation. As mentioned above, in the MiniBooNE analysis,
ejected nucleons are not detected and the QE cross section is defined
as the one for processes in which only a muon is detected in the final
state. The MiniBooNE analysis of the data corrects (through a Monte
Carlo estimate) for events, where in the neutrino interaction a pion
is produced, but it escapes detection because it is reabsorbed in the
nucleus, leading to multinucleon emission. However, in
~\cite{Martini:2009uj, Martini:2010ex} it is pointed out that 2p2h or
3p3h mechanisms are susceptible to produce an apparent increase in the
``QE'' cross section, since those events will give rise to only one
muon to be detected, and the MiniBooNE analysis does not correct for
them. Within the scheme followed in Ref.~\cite{Benhar:2010nx}, the
occurrence of 2p2h final states is described by the continuum part of
the spectral function, arising from nucleon-nucleon correlations, and
there, this contribution is found to be quite small (less than 10\% of
the integrated spectrum). This is not surprising, since our QE results
(dashed line) in the left panel of Fig.\ref{fig:sQE} contain also this
contribution\footnote{The CCQE cross sections calculated in
Ref.~\cite{Nieves:2004wx}, were obtained using both particle and hole
dressed propagators, determined from a realistic in medium nucleon
selfenergy~\cite{FernandezdeCordoba:1991wf}, and thus account for the
spectral function effects considered in \cite{Benhar:2010nx}. }, and
as we mentioned, we underestimate the data. However, the 2p2h
contribution considered in \cite{Benhar:2010nx} is far from being
complete\footnote{In fact, the spectral function model taken as a {\it
paradigm} in the discussion of Ref.~\cite{Benhar:2010nx}, though
successful to account for the QE electron--nucleus scattering, at
intermediate energies, badly fails to describe both the dip and the
$\Delta-$regions, as can be appreciated for instance in Figs. 5-8 of
Ref.~\cite{Benhar:2005dj} or in the Fig. 1 of
Ref.~\cite{Benhar:2010nx}. This is because the lack of a proper model
to account for the absorption of the virtual photon for two or three
nucleons in that model.} and it corresponds only to the many body
diagram depicted in Fig.~\ref{fig:PNterm}. Here, we compute all the
contributions contained in the generic diagrams of
Figs.~\ref{fig:2ph}, \ref{fig:2ph2bubbles} and \ref{fig:rho-vt}, as it
was previously done in Ref.~\cite{Gil:1997bm} for electron scattering,
obtained from a realistic model for the weak pion production off the
nucleon. When these latter contributions are added to the QE
prediction of Ref.~\cite{Nieves:2004wx}, we obtain the solid green
line in the left plot of Fig.\ref{fig:sQE} in a better agreement with
the MiniBooNE data.

As commented before, these multinucleon knockout events 
are likely part of  the CC``QE'' cross section measured by MiniBooNE,
and that naturally explains the failure of the scheme of
Refs.~\cite{Benhar:2005dj, Benhar:2010nx}.

Coming back to the left plot of Fig.\ref{fig:sQE}, there we also
display the band of theoretical uncertainties affecting our
results. To estimate this band, we have summed in quadratures a 15\%
relative error in our results with the error induced by the
uncertainties on the parametrization of the $C_5^A(q^2)$ form factor
used here (set IV in Table I of Ref.~\cite{Hernandez:2010bx}). As
discussed in Ref.~\cite{Valverde:2006zn} for the CCQE case, 15\% is a
conservative ansatz to account for the errors, in total and
differential inclusive cross sections, induced by the uncertainties affecting the
nuclear corrections included in our model.  Once, our theoretical
uncertainties are taken into account, we find a reasonable agreement
with the MiniBooNE data. We would like to stress that we have not
fitted here any parameter, and that we have just extended our previous
work on electron-scattering of Ref.~\cite{Nieves:2004wx} to the study
of CCQE cross sections.

In Fig.~\ref{fig:sQE}, we also show the results of Martini and
collaborators (blue dash-dotted line), taken from the {\it QE+np-nh RPA}
curves of Fig. 5 of Ref.~\cite{Martini:2010ex}, which nicely fall
within our band of theoretical predictions. Details of the model used
by M. Martini and collaborators can be found in
Ref.~\cite{Martini:2009uj}. The evaluation of the nuclear response
induced by these 2p2h mechanisms carried out in
Ref.~\cite{Martini:2009uj} is approximated, as explained there.  
The contributions in \cite{Martini:2009uj} that can be
cast as a $\Delta-$selfenergy diagram should be quite similar to those
obtained here in Subsect.~\ref{sec:delta}, since in both cases the
results of Ref.~\cite{Oset:1987re} for the
$\Delta-$selfenergy are  used. However, some other contributions
included here are, either not considered or not properly taken into
account in \cite{Martini:2009uj}. For example, we believe that none of
diagrams of Fig.~\ref{fig:2ph2bubbles} or those in Fig.~\ref{fig:2ph}
involving the $CT$, $PP$ and $PF$ vertices of Fig.~\ref{fig:diagramas}
have been considered in the work of Martini and
collaborators. Moreover, the $NP-NP$, $CNP-CNP$, $NP-CNP$, $NP-\Delta
P$, $NP-C\Delta P$, $CNP-\Delta P$, $CNP-C\Delta P$ and $\Delta
P-C\Delta P$ diagrams implicit in Fig.~\ref{fig:2ph}, are not directly
evaluated in \cite{Martini:2009uj}, but instead, an
indirect estimate is given for them by relating their contribution to
some absorptive part of the $p-$wave pion-nucleus optical
potential. Given all this, we find remarkable the agreement exhibited
in Fig.~\ref{fig:sQE} between our results and those previously
published in Refs.~\cite{Martini:2009uj,Martini:2010ex}.

On the other hand, we see that in our calculation the relative 
contribution of the 2p2h mechanisms with respect to the QE cross
section, is quite similar for both neutrino and antineutrino induced
processes. Thus, in what respect to this issue, our results do not
support the claims of Ref.~\cite{Martini:2010ex} on a minor role of
the  2p2h mechanisms in the antineutrino mode.

We should mention that the MiniBooNE collaboration has also
published the flux-integrated CC''QE'' double differential cross section
$d^2\sigma/dE_\mu d \cos\theta_\mu $ in bins of muon energy $E_\mu$
and cosine of the muon scattering angle with respect to the incoming
neutrino direction. We must refrain to compare with these valuable
data. The reason is that the MiniBooNE flux remains sizeable up to 
neutrino energies too high to make meaningful the predictions of the
model presented here. Indeed, neutrino energies of 
1 or 1.2 GeV at most, is the clear upper limit of validity of our
predictions. The fraction of the MiniBooNE flux above 1.2 GeV is still
larger than 17\%, and this together with the fact that  the
cross sections grows with the energy has prevented us to do the
comparison. We are working in extending our model to higher energies,
but this is far from being a trivial task. New channels (2p2h1$\pi$ or
two pion production, \dots) and higher resonances must be
incorporated into the model, besides of smoothly getting rid of the RPA
and other nuclear effects, which importance diminishes when the
transferred energy increases. 

\begin{figure}
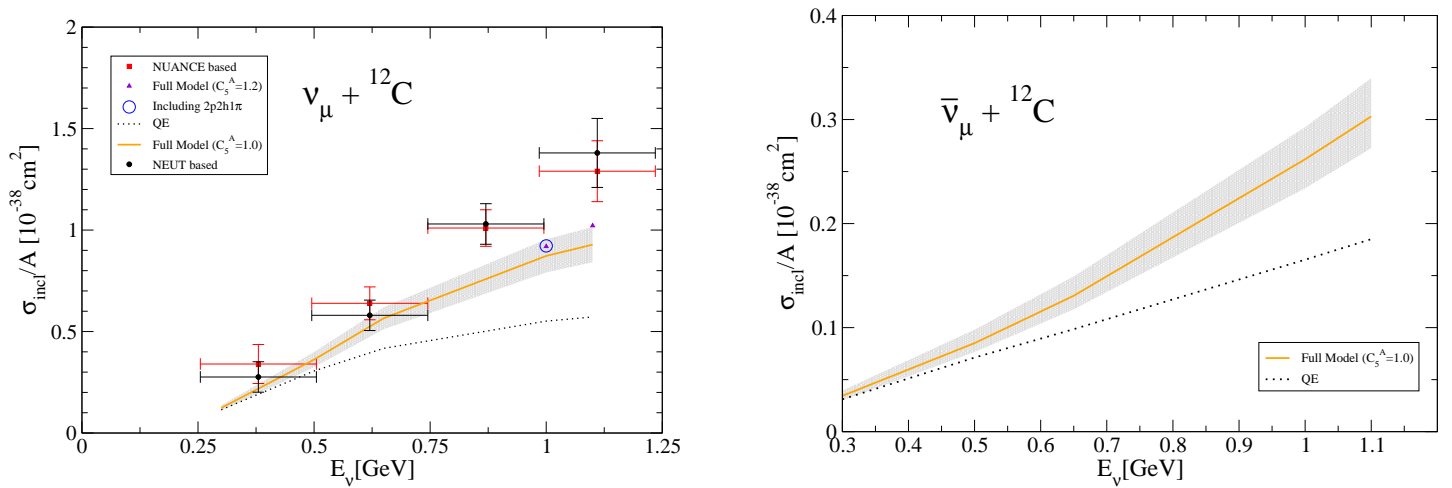

\makebox[0pt]{\includegraphics[width=0.5\textwidth]{Sigma_Total_Inclusiva_perNucleonWithErrorBands3.eps}\hspace{1cm}
\includegraphics[width=0.5\textwidth]{SigmaTotalInclusiva_pernucleonWitherrorbands.eps}}
\caption{Left: Data points stand for the SciBooNE neutrino CC
inclusive interaction cross section per
nucleon~\cite{Nakajima:2010fp}. We also show our QE and full model
results, and in this latter case the theoretical uncertainty band is
also displayed. At 1 GeV, we depict the full model cross sections
obtained when the GTR value of 1.2 for $C_5^A(0)$ is used instead of 1
(violet triangle), and when some 2p2h$1\pi$ contributions (blue empty
circle), neglected in the present work, are taken into account (see
text for some more details).  We also give at 1.1 GeV the total cross section
obtained with $C_5^A(0)= 1.2$. Right: QE and full model
predicted antineutrino CC inclusive cross section per nucleon, as a
function of the antineutrino energy.}
\label{fig:stot2}
\end{figure}
To end this section in Fig.~\ref{fig:stot2} we show total and QE
inclusive cross sections for both neutrino and antineutrino modes. In
the neutrino case, we compare our results with the recent data
published by the SciBooNE collaboration. We display SciBooNE data-sets
based on NEUT and NUANCE Monte Carlo event generators. We find a
reasonable description, taking into account experimental and
theoretical uncertainties, up to neutrino energies around 1 GeV. At
larger energies, we underestimate the cross section, as anticipated
above. For instance, we see how some 2p2h1$\pi$ contributions
neglected in our model, become relatively important at $E_\nu= 1$
GeV. More specifically, the empty circle is obtained when the
$\Delta-$resonance contribution to the imaginary part of $U$ is kept
in the evaluation of the imaginary part of the $\pi-$ and
$\rho-$selfenergies in Eqs.~(\ref{eq:dy-2p2h}) and
(\ref{eq:rho-self}). There are some other $W^+NN\to NN\pi$ mechanisms
which should be taken into account, as well as the contribution of
higher resonances~\cite{Lalakulich:2006sw}.  Though small, also
kaon~\cite{RafiAlam:2010kf}, hyperon~\cite{Singh:2006xp} and two
pion~\cite{Hernandez:2007ej} production channels should be considered 
to end up with a robust theoretical model above 1 GeV.

\section{Conclusions} 
\label{sec:conc}

We have developed a model for the study of weak CC induced
nuclear reactions at intermediate energies of interest for current
and future neutrino oscillation experiments. This model is an 
extension of the work of Ref.~\cite{Nieves:2004wx} that
analyzed the QE contribution to the inclusive neutrino scattering on
nuclei. The model is based on a systematic many body expansion of the
gauge boson absorption modes that includes one, two and even three
body mechanisms, as well as the excitation of $\Delta$ isobars. The
whole scheme has no free parameters, besides those previously adjusted
to the weak pion production off the nucleon cross sections in the
deuteron, since all nuclear effects were set up in previous studies of
photon, electron and pion interactions with nuclei.

We have discussed at length the recent CCQE MiniBooNE cross section
data.  To understand these measurements, it turns out to be essential the
consideration of mechanisms where the $W-$boson is absorbed by
two or more nucleons without producing pions, as first suggested by
M. Martini and collaborators~\cite{Martini:2009uj}.  Our evaluation of
these pionless multinucleon emission contributions to the cross
section is fully microscopical and it contains terms, which were
either not considered or only  approximately taken into account in
\cite{Martini:2009uj}. We end up with a reasonable description of the
neutrino CC''QE'' MiniBooNE and total inclusive SciBooNE cross section
data up to neutrino energies of around 1 GeV.

Our results do not support  the  incompatibility among
neutrino and electron-nucleus inclusive data claimed in 
\cite{Benhar:2010nx}, since our neutrino model is just a natural
extension of that developed in Refs.~\cite{Gil:1997bm} and
\cite{Carrasco:1989vq} to study electron-- and photo-nuclear inclusive
reactions. Indeed, we believe that the origin of the problem can be
traced back to the difficulties of the spectral function model
advocated in \cite{Benhar:2010nx} to properly describe the {\it dip}
and $\Delta-$ regions for electron scattering, together with the
mismatch existing in the definition of the quasielastic contribution
between the theory and the experimental neutrino communities.

The recent CC MiniBooNE and SciBooNE inclusive data sets provide very
valuable information to distinguish among different models. This will
definitely help to unravel the details about the modification of the
CC weak current properties   inside of the nucleus, and will set up
the basis to construct a robust theoretical framework where all
electroweak nuclear reactions at intermediate energies could be
studied. This is in turn of  special relevance to better understand
the systematic errors affecting present (MiniBooNE \& T2K) and coming 
neutrino oscillation experiments involving neutrinos with energies
below 1 GeV.   Future antineutrino data, 
similar to the CCQE MiniBooNE measurements
of total and differential neutrino cross sections, will further
constraint any theory. 

We think the microscopical model presented here, which
extends that  of Ref.~\cite{Nieves:2004wx} beyond the QE region,
constitutes a first step towards this goal. The model
should be extended still to higher energies, which would make
possible the comparison of its predictions to the MiniBooNE
differential cross section data. This is a non trivial task, but it
will allow for a better knowledge of the axial currents both for
hadrons and nuclei.

\begin{acknowledgments}
This research was supported by DGI and FEDER funds, under contracts
  FIS2008-01143/FIS, FIS2006-03438, and the Spanish Consolider-Ingenio
  2010 Programme CPAN (CSD2007-00042), by Generalitat Valenciana
  contract PROMETEO/2009/0090 and by the EU HadronPhysics2 project,
  grant agreement n. 227431.  I.R.S. acknowledges support from the
  Ministerio de Educaci\'on.
\end{acknowledgments}


\end{document}